\newif\iffigs\figstrue

\documentclass[12pt]{article}
\usepackage{amstex,epsf,amssymb}
\DeclareFontFamily{U}{rsf}{}
\DeclareFontShape{U}{rsf}{m}{n}{
  <5> <6> rsfs5 <7> <8> <9> rsfs7 <10-> rsfs10}{}
\DeclareMathAlphabet\Scr{U}{rsf}{m}{n}

\def\pplogo{\vbox{\kern-\headheight\kern -15pt
\halign{##&##\hfil\cr&{%\sc
\ppnumber}\cr\rule{0pt}{2.5ex}&\ppdate\cr}
}}
\makeatletter
\def\ps@firstpage{\ps@empty \def\@oddhead{\hss\pplogo}%
  \let\@evenhead\@oddhead % in case an article starts on a left-hand page
}
\def\maketitle{\par
 \begingroup
 \def\thefootnote{\fnsymbol{footnote}}
 \def\@makefnmark{\hbox
 to 0pt{$^{\@thefnmark}$\hss}}
 \if@twocolumn
 \twocolumn[\@maketitle]
 \else \newpage
 \global\@topnum\z@ \@maketitle \fi\thispagestyle{firstpage}\@thanks
 \endgroup
 \setcounter{footnote}{0}
 \let\maketitle\relax
 \let\@maketitle\relax
 \gdef\@thanks{}\gdef\@author{}\gdef\@title{}\let\thanks\relax}
\makeatother

\def\C{{\mathbb C}}
\def\P{{\mathbb P}}
\def\Q{{\mathbb Q}}

\def\Z{{\mathbb Z}}

\def\Hom{\operatorname{Hom}}

\def\SO{\operatorname{SO}}

\def\GU{\operatorname{U{}}}
\def\Sp{\operatorname{Sp}}

\def\Spin{\operatorname{Spin}}
\def\so{\operatorname{\mathfrak{so}}}
\def\su{\operatorname{\mathfrak{su}}}
\def\gu{\operatorname{\mathfrak{u}}}
\def\sp{\operatorname{\mathfrak{sp}}}

\def\CY{Calabi--Yau}

\def\cM{{\Scr M}}

\def\cD{{\Scr D}}

\def\cMc{{\hfuzz=100cm\hbox to 0pt{$\;\overline{\phantom{X}}$}\cM}}
\def\barcD{{\hfuzz=100cm\hbox to 0pt{$\;\overline{\phantom{X}}$}\cD}}
\def\ff#1#2{{\textstyle\frac{#1}{#2}}}

\def\spnh{\Spin(32)/\Z_2}
\def\mbf#1{\mbox{\mathversion{bold}$#1$}}

\begin{document}
\setcounter{page}0
\title{\LARGE Point-like Instantons and the\\ $\spnh$ Heterotic String\\[10mm]}
\author{
Paul S. Aspinwall\\[0.7cm]
\normalsize Dept.~of Physics and Astronomy,\\
\normalsize Rutgers University,\\
\normalsize Piscataway, NJ 08855\\[10mm]
}
\def\ppnumber{\vbox{\baselineskip14pt\hbox{RU-96-113}
\hbox{hep-th/9612108}}}
\def\ppdate{December 1996} \date{}

{\hfuzz=10cm\maketitle}

\def\Large{\large}
\def\LARGE{\large\bf}

\vskip 1cm

\begin{abstract}
We consider heterotic string theories compactified on a K3 surface
which lead to an unbroken perturbative gauge group of $\spnh$. All solutions
obtained are combinations of two types of point-like instanton --- one
``simple type'' as discovered by Witten and a new type associated to
the ``generalized second Stiefel-Whitney class'' as introduced by
Berkooz et al. The new type of instanton is associated to an
enhancement of the gauge symmetry by $\Sp(4)$ and the addition of a
massless tensor supermultiplet. It is shown that if four simple
instantons coalesce at an orbifold point in the K3 surface then a
massless tensor field appears which may be used to interpolate between
the two types of instanton. 
By allowing various combinations of point-like instantons to coalesce,
large gauge groups (e.g., rank 128) with many massless tensor
supermultiplets result. 
The analysis is done in terms of F-theory.
\end{abstract}

\vfil\break

%%%%%%%%%%%%%%%%%%%%%%%%%%%%%%%%%%%%%%%%%%%%%%%%%%%%%%%%%%%%%%%%

\section{Introduction}

There has recently been considerable progress in the understanding of
the nonperturbative physics of string compactification. A
fairly realistic model which would be very nice to understand would be
the heterotic string compactified on a \CY\ threefold as this leads to
an $N=1$ theory in four dimensions. Here we will deal with the
more modest model of a heterotic string compactified on a K3
surface to yield an $N=1$ theory in six dimensions.

Starting with the work of \cite{KV:N=2,FHSV:N=2} it was realized that
the structure of the heterotic string on a K3 surface could be related
to the geometry of a \CY\ threefold. In particular that the type IIA
string on this \CY\ space was dual to the heterotic string
compactified on a product of a K3 surface and a 2-torus. Recall that
the heterotic string requires a bundle structure for its
compactification and so this product of K3 and a torus also comes
equipped with a bundle.

By a process known as F-theory \cite{Vafa:F,MV:F,MV:F2} one can
analyze only the parts of the \CY\ threefold, $X$, that are relevent to the
K3 part of the compactification and ignore the 2-torus part. In order
to do this, $X$ must be in the form of an elliptic fibration with section
$p:X\to\Theta$, where $\Theta$ is a complex surface. For a 
precise statement of this see \cite{me:lK3}. This may be viewed in two
ways. Firstly one may take the area of the $T^2$ to be large and
switch off any Wilson lines around it and watch how $X$
degenerates. Alternatively one may perform a fibre-wise mirror map and
replace the type IIA theory with a type IIB string compactified on
$\Theta$, where points in $\Theta$ corresponding to ``bad fibres'' are
identified with D-branes embedded in the space.
Either way, F-theory promises to yield a fairly complete
understanding of the entire moduli space of heterotic strings on a K3
surface.\footnote{Although the moduli associated with the R-R sector
in the type IIA string may be troublesome.}

Since we are able to probe the moduli space so well, we should ask where
the interesting points might be. An obvious place to look is where the
underlying K3 surface itself degenerates to an orbifold. This,
afterall, has been where the interesting physics lives when one
compactifies a type IIA or IIB string on a K3 surface
\cite{W:dyn,me:enhg,W:dyn2}. It turns out that a simpler question to
answer concerns when the bundle data on the K3 surface
degenerates. This has no analogue for the type II string on the K3
surface.

The most obvious type of degeneration of a bundle is that of
the ``point-like instanton''. That is, where the curvature of the bundle
is concentrated in an infinitesimal region of the base space. The
study of such objects in heterotic string theory began with Witten's
paper \cite{W:small-i}. Here it was argued that for the $\spnh$
heterotic string on a smooth K3 surface a point-like instanton induces
massless vector multiplets enhancing the gauge symmetry by
$\sp(1)$.\footnote{Unless we really need to, we will speak in terms of the
algebra, rather than the group, of the gauge symmetry. This will allow
us to ignore a lot of awkward $\Z_2$ factors.} This was suggested on
general grounds, from the quaternionic nature of the moduli space of
hypermultiplets and pictured in terms of Dirichlet 5-branes from the
dual type I string theory. When $k$ point-like instantons coalesce at
the same point in the K3 surface it was argued that the gauge symmetry
is enhanced by $\sp(k)$.

In the context of point-like instantons, the $E_8\times E_8$ heterotic
string appears, at first sight, to be a quite different animal. From
duality to M-theory \cite{HW:E8M}, it was argued in \cite{SW:6d} that
point-like instantons induce peculiar ``tensionless strings'' and new
moduli in tensor supermultiplets which 
allow one to move off in a new direction in moduli space corresponding
to a new massless tensor multiplet in the theory.

This is far from the end of the story for point-like instantons
however. What if the underlying K3 surface is singular and a
point-like instanton sits right at the singularity? This may well
provide new behaviour. The purpose of this paper is to give an example
of such an instanton in the context of the $\spnh$ heterotic string
and explore some of its rich properties.

There are two approaches to nonperturbative analysis of the heterotic
string by duality. One method, which we use here, is F-theory, which
may be viewed as finding a type II dual. The rival method is that of
using duality to the type I string, in particular by using orientifold
methods \cite{GP:open}. One should note that one may directly relate
these two approaches to each other of course \cite{Sen:ort}. The
orientifold approach has proven very powerful in its ability to find
explicit spectra for given models --- see, for example,
\cite{DP:ort,GJ:ort}. Indeed, the subject of massless tensors
associated to the $\spnh$ heterotic string was analyzed in this
context in \cite{Pol:ortt}.

We wish to attempt to completely classify heterotic string theories on a K3
surface which contain the original ten-dimensional $\spnh$ as part of
their unbroken gauge symmetry. This
will lead us to the new instanton.
In general one should expect the F-theory approach to give a much
better coverage of the moduli space of theories than the orientifold
approach. This is because the orientifold approach necessarily focuses
on points in the moduli space corresponding the theories which are the
global quotient of some other theory. One may also probe an
infinitesimal region around this point by ``twisted marginal
operators''. 
F-theory on the other hand phrases questions in terms of
elliptic fibrations. Since any smooth deformation of an elliptic
threefold is also an elliptic threefold one might at first think one
can probe the entire moduli space of a given theory. While this is
almost true, current understanding of F-theory only treats enhanced
gauge symmetry from the point of view of degenerate fibres. There is
another potential contribution from the ``Mordell-Weil group''. This
arises when the fibration has an infinite number of sections. We will
ignore this latter possibility.

It is not clear whether or not orientifold techniques could reproduce
the results in this paper but it would be an interesting question to
answer.\footnote{I thank C.~Johnson for a correspondence on this question.}

We will present the classical geometry of this new instanton in
section \ref{s:bun} and relate it to the ``generalized second
Stiefel-Whitney class'' as introduced by Berkooz, Leigh, Polchinski, Schwarz,
Seiberg, and Witten \cite{BLPSSW:so32}. This will allow us to build
our new ``hidden obstructer'' point-like instanton in section
\ref{s:ghol} which has nonzero generalized second Stiefel-Whitney
class but manages to not break any of the primordeal $\spnh$ gauge
group.

In section \ref{s:F} we build the F-theory picture of the new
instanton which allows us to determine its nonperturabtive
physics. The main result is the appearance of an $\sp(4)$ enhanced
gauge symmetry and a massless tensor. In section \ref{s:eq} the
F-theory picture is tied to known results about the $E_8\times E_8$
heterotic string and to the Gimon-Polchinski models.

In section \ref{s:ph} we show how to transform our new instantons into
Witten's simple instantons and {\em vice versa}. This will also show
that four simple instantons coaleced at an orbifold point in the K3
surface induce a massless tensor.

In section \ref{s:co} we tackle the question of what happens when
the two types of instantons hit each other. Large spectra of gauge
symmetries and hypermultiplets appear. Finally we include an appendix
which reviews some properties of elliptic threefolds that we require.

%%%%%%%%%%%%%%%%%%%%%%%%%%%%%%%%%%%%%%%%%%%%%%%%%%%%%%%%%%%%%%%%%%%

\section{Bundles on K3 and the Kummer Lattice}  \label{s:bun}

Let us review the notion of a {\em generalized second Stiefel-Whitney
class\/} following the work of Berkooz et al \cite{BLPSSW:so32}.
Consider a smooth $G$-bundle, $E$, on a smooth K3 surface, $S$. How
can we express the topology of this bundle? Consider a 2-sphere, $C$,
within $S$ with a curve, $\gamma$, around its equator. An element,
$g_\gamma\in G$, of the holonomy of $E$ may be found by parallel
transport around this curve.

View $C$ as the union of its northern hemisphere, $C_N$, with its
southern hemisphere, $C_S$. From the curvature, $F$, of $E$ we may
then determine
\begin{equation}
\begin{split}
  g_\gamma &= \exp\left(\int_{C_N} iF\right)\\
  &= \exp\left(-\int_{C_S} iF\right).
\end{split}
\end{equation}
Thus
\begin{equation}
  \int_C F = 2\pi n,   \label{eq:c1}
\end{equation}
for some integer, $n$. Thus $\ff1{2\pi}F$ appears as an element of
$H^2(S,\Z)$. This quantity will depend on the topology of $E$ but it
may be that different values of $n$ specify the same topological
class. To see how this works, consider the transition functions around
$\gamma$ from the northern hemisphere to the southern hemisphere as a
map from $\gamma$ into $G$. In order that $E$ be homotopically
nontrivial we require that the image of $\gamma$ lie in a nontrivial
element of $\pi_1(G)$. We may apply this construction to every homology
2-cycle within $S$.

We arrive at the result that a natural topological invariant of a
$G$-bundle on $S$ is given by a homomorphism from $H_2(S,\Z)$ to
$\pi_1(G)$. If $\pi_1(S)$ is trivial then the universal coefficients
theorem \cite{BT:} says that this group of homomorphisms is isomorphic to
$H^2(S,\pi_1(G))$.

There are two very familiar examples of this invariant. First if $E$ is
the principle bundle of a holomorphic vector bundle then
$G\cong\GU(r)$, for some $r$. Since $\pi_1(\GU(r))\cong\Z$ we have our
invariant is simply an element of $H^2(S,\Z)$. This is the {\em first
Chern class}, $c_1(E)$. If $E$ is the principle bundle of a real vector bundle
then $G\cong\SO(r)$. Since $\pi_1(SO(r))\cong\Z_2$ we have an object
in $H^2(S,\Z_2)$. This is the {\em second Stiefel-Whitney class},
$w_2(E)$. 

We are interested in the case $G\cong\spnh$. Clearly
$\pi_1(\spnh)\cong\Z_2$ and so we are in a situation analogous to the
second Stiefel-Whitney class. Following \cite{BLPSSW:so32} we denote
this $\tilde w_2\in H^2(S,\Z_2)$ and consider it to be a
``generalized second Stiefel-Whitney class''.

It will be convenient to represent $\tilde w_2$ as a 2-cycle rather
than a 2-cocycle. Dual to $H^2(S,\Z)$ is $H_2(S,\Z)$ in the usual
way. We may then take $H_2(S,\Z)$ to be dual to itself by Poincar\'e
duality. Thus we may take the dual of the dual of an element of
$H^2(S,\Z_2)$ as an element of $H_2(S,\Z_2)$. When seen this way,
$\tilde w_2$, as an element of $\Hom(H_2(S),\Z_2)$, may be viewed as
\begin{equation}
  \tilde w_2:C\to\#(\tilde w_2\cap C)\pmod 2,
\end{equation}
where ``$\#$'' represents the intersection number. We will simply use a
dot to represent this natural inner product in $H_2(S,\Z)$ from now
on.

We will be particularly interested in the case where the K3 surface,
$S$, is a {\em Kummer Surface}. That is, when it has been obtained as
the blow-up of the orbifold $T^4/\Z_2$ in the usual way. The Kummer
surface gives a natural set of elements in $H_2(S,\Z)$. These are
\begin{enumerate}
\item The image of the six 2-cycles in the $T^4$ under the quotient
map.
\item The sixteen 2-spheres that appear as the exceptional divisors under
blowing up.
\end{enumerate}
Although these 22 elements may be used as a basis for $H_2(S,\Q)$,
they are not correctly normalized to form a basis for
$H_2(S,\Z)$. That is, they generate only a finite-index sublattice of 
$H_2(S,\Z)$. This sublattice is called the {\em Kummer Lattice}.
$H_2(S,\Z)$ is even self-dual, whereas
the matrix of inner products on the generators of the Kummer Lattice
has determinant not equal to one.

Let us use $C_i, i=1\ldots16$, to denote the sixteen exceptional
divisors. Since $C_i.C_i=-2$ by the usual arguments (see, for example,
\cite{me:lK3}), then $C_i/n$ cannot be an element of $H_2(S,\Z)$ for
any integer, $n>1$. Having said that, certain sums of $C_i's$ will be
multiples of elements in $H_2(S,\Z)$. This partially accounts for why
$\{C_i\}$ are not good generators for $H_2(S,\Z)$. Let
\begin{equation}
  D = \sum_{i=1}^{16}\xi_iC_i,
\end{equation}
where $\xi_i$ is either 0 or 1. One may then show that $D$ will be
{\em twice\/} an element of $H^2(S,\Z)$ if the following is true.
The 16 exceptional divisors come from the 16 fixed points of the
$\Z_2$ action on $T^4$. The latter sixteen points naturally form the
vertices of a hypercube.
Consider every two-dimensional face of the hypercube. Each such face
will contain four $C_i$'s and thus can be associated with four
$\xi_i$'s. The sum of these four $\xi_i$'s must be an even number. 

A simple solution is to set all $\xi_i$ to zero, which is trivial, or
to set all $\xi_i$ to one. The other possibilities correspond to
having eight $\xi_i$'s equal to zero and eight $\xi_i$'s equal to one
in suitable combinations.

In section 4.1 of \cite{BLPSSW:so32} a picture of an instanton with
$\tilde w_2\neq0$ 
was given locally for an open neighbourhood of one of the exceptional
divisors. This was given in terms of the curvature of the
bundle which could be given compact support near the exceptional
divisor. 
One may try to treat a $\spnh$-bundle as if it were
a $\Spin(32)$-bundle simply by viewing the transition functions as
elements of $\Spin(32)$ rather than $\spnh$. As such we may
try to build a bundle in the vector representation. Let the curvature
of this resulting bundle be $F_{\mathbf{32}}$. One may then show that
\begin{equation}
  \int_C \ff1{2\pi}F_{\mathbf{32}} = \ff12(\tilde w_2.C) + n,
\end{equation}
for some integer, $n$. Thus $\tilde w_2$ can violate the quantization
condition (\ref{eq:c1}) and obstruct the existence of a vector
representation --- just as $w_2$ obstructs a spin structure. If $C_i$
is the exceptional divisor in question then the instanton of
\cite{BLPSSW:so32} satisfies
\begin{equation}
\int_{C_i} \ff1{2\pi}F_{\mathbf{32}} = \ff12,   \label{eq:s32}
\end{equation}
and thus obstructs a vector structure over $C_i$. We will call this
instanton a ``$C_i$-obstructer''. A $C_i$-obstructer can be seen to
satisfy $\tilde w_2.C_i=1$.

Since the curvature of a $C_i$-obstructer is meant to arise from local
geometry, $\tilde w_2$ should be proportional to $C_i$. Since
$C_i.C_i=-2$, we have
\begin{equation}
  \tilde w_2 = \ff12C_i.
\end{equation}

Now let us fit the local $C_i$-obstructer picture into the global
geometry of the K3 surface, $S$. It is clear that a single obstructer
is not a valid configuration since $\tilde w_2$ does not lie in
integral homology. We may consider a situation where we place a
single obstructer over more than one exceptional divisor. Now, if our
set of exceptional divisors satisfies the Kummer lattice condition
above then we are in business. The solution considered in \cite{BLPSSW:so32}
was to put an obstructer at all sixteen sites and so
\begin{equation}
  \tilde w_2 = \ff12\sum_{i=1}^{16}C_i,
\end{equation}
which is in integral homology.

%%%%%%%%%%%%%%%%%%%%%%%%%%%%%%%%%%%%%%%%%%%%%%%%%%%%%%%%%%%%%%%%%%%%%%%

\section{Unbroken Gauge Symmetry and Global Holonomy}
    \label{s:ghol}

When considering a compactification of a heterotic string on a bundle,
$E\to B$, an important piece of information about $E$ is its {\em global
holonomy}, $H$. Let $G_0$ be the ``primordial'' gauge group, i.e.,
$E_8\times E_8$ or $\spnh$, of the heterotic string in ten
dimensions. When compactified on $E$, this will be broken to the {\em
centralizer\/} of $H\subset G_0$. That is, any element of $G_0$ which
commutes with all of $H$ will remain a symmetry after
compactification.

There are two contributions to the global holonomy group, $H$. Firstly
there is the local holonomy generated by the curvature of
$E$. Secondly there is the contribution from non-contractable loops
from $\pi_1(B)$ of the base space of $E$.

We will be most interested in the case where the global holonomy group
is trivial and thus all of the primordial gauge symmetry remains in
the lower-dimensional compactified theory. We need to make both the
local holonomy and the contribution from $\pi_1$ trivial.

We know from the work of \cite{W:small-i} how to make the local
holonomy trivial. Since this comes from the curvature of the bundle,
we need to squeeze all of the region of the nonzero curvature into
points over the base space. This limit is called ``point-like
instantons''. Of course, we haven't really justified that the paths
which happen to exactly pass through the point where a small
instanton lives don't pick up holonomy but the evidence is
considerable \cite{W:small-i,AG:sp32} that string theory really does
allow one to ignore such paths.

Once we have shrunken all instantons down to zero size, we need only
worry about non-contractable loops breaking the gauge group. Actually
we should consider loops that are non-contractable {\em after\/} the
points within $B$ where the point-like instantons live have been
removed since we are required to ignore paths which pass through such
points. 

Let us consider the case where $B$ is a K3 surface. Since a K3 surface
is simply connected, we need only worry about non-contractable loops
produced by removing the locations of point-like instantons. If the
instanton happens to sit at a smooth point inside the K3 surface then
the open neighbourhood of the instanton, minus the point where it sits,
may be retracted onto $S^3$ --- which is simply connected. Thus, a
point-like instanton at a smooth point in a K3 surface breaks non of
the primordial gauge group. For the $\spnh$ heterotic string, such
point-like instantons are precisely the ones discovered by Witten
\cite{W:small-i}. We will denote such point-like instantons ``simple''.

In the case that the K3 surface is a Kummer surface at an orbifold
limit, we have a singularity locally of the form $\C^2/\Z_2$. If the
instanton happens to be sat right on this singular point, then the
neighbourhood retracts onto the lens space $S^3/\Z_2$. Since
$\pi_1(S^3/\Z_2)$ equals $\Z_2$ we now have the possibility that the
point-like instanton breaks part of the primordial gauge symmetry.

As shown in \cite{BLPSSW:so32} this breaking of the gauge symmetry by
$\pi_1$ effects is intimately connected to $\tilde w_2$ of the
instanton. Let us review this fact. Consider blowing up the orbifold
slightly so that we have an exception 2-sphere, $C_i$, in a small open
neighbourhood of the K3 surface. We may also put the lens space
$S^3/\Z_2$ in this open neighbourhood, surrounding the 2-sphere. We
show this in figure \ref{fig:Lens}.

\iffigs
\begin{figure}[t]
  \centerline{\epsfxsize=7cm\epsfbox{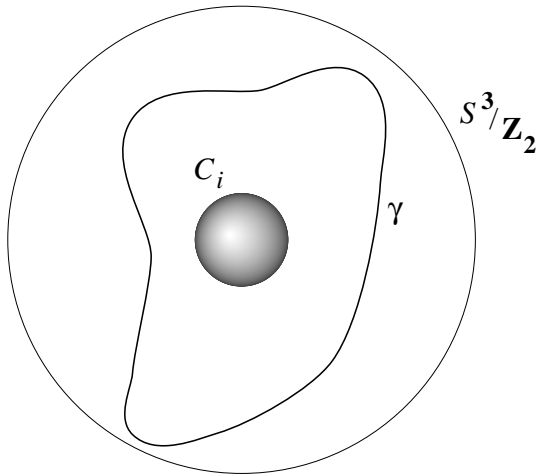}}
  \caption{Lens space around an exceptional divisor.}
  \label{fig:Lens}
\end{figure}
\fi

Now, the lens space, $L$, may be viewed as an $S^1$-bundle over the
2-sphere, $C_i$. We may then use the Leray spectral sequence (see
example 15.15 in \cite{BT:}) to write the cohomology of $L$ in terms of
that of $C_i$. The important point is that there is an isomorphism
\begin{equation}
  \phi:H^2(C_i,\Z_2) \cong H^2(L,\Z_2)\cong\Z_2.
\end{equation}
This maps the topological class of bundles over $C_i$ as measured by
$\tilde w_2$ into the class of bundles over $L$ given by $\phi(\tilde
w_2)$. The generator of $H^2(L,\Z_2)$ may be associated to the generator of
$\pi_1(L)$. This follows from the universal coefficients theorem
\cite{BT:} and the fact that $H^2(L,\Z)$ is pure torsion. Let us call
this latter generator, $\gamma$. We show $\gamma$ as a
non-contractable loop in figure \ref{fig:Lens}.

If $\tilde w_2.C_i=1$ then the bundle is nontrivial. The only way the
bundle on $L$ may be nontrivial is if the holonomy element generated
by $\gamma$ is nontrivial. Thus the global holonomy of the instanton
is precisely measured by $\tilde w_2$. 

As discussed in \cite{BLPSSW:so32}, the $\Z_2$ subgroup of $\spnh$
generated by $\gamma$ is unique, up to endomorphisms. It is not the
central $\Z_2$ in $\spnh$ and actually 
breaks the primordial gauge group down to $\GU(16)/\Z_2$. That is,
{\em a point-like instanton in the form of a $C_i$-obstructer breaks
$\spnh$ to $\GU(16)/\Z_2$.}

This isn't what we want however. We want to see if we can leave the
entire $\spnh$ unbroken. There is a very simple way of producing a
bundle with $\tilde w_2\neq0$ and yet keeping the primordial $\spnh$
intact. Consider the case where
\begin{equation}
  \tilde w_2 = C_i.
\end{equation}
Now we have $\tilde w_2(C_i)=C_i.C_i\pmod2=0$. That is, the bundle
over $C_i$, and hence $L$, is now topologically trivial. The
holonomy around $\gamma$ will be trivial and so $\spnh$ remains
unbroken when $C_i$ is blown down to a point. Let us call this the
hidden-$C_i$-obstructer. 

At first sight it looks like we have constructed something rather
trivial but consider the case where we have precisely one curve,
$C_i$, in the K3 surface over which we put the hidden obstructer. Then
as $C_i$ is not twice an element of $H^2(S,\Z_2)$ we really do have a
nontrivial value for $\tilde w_2$. Following our discussion of the Kummer
lattice in the previous section, the situation for hidden
obstructers is somewhat the opposite as for the previously discussed
non-hidden obstructers: 
\begin{itemize}
  \item Non-hidden obstructers must appear in multiples of eight so
that the associated curves add up to twice an element of the Picard
lattice.
  \item Hidden obstructers must {\em not\/} appear in such multiples
of eight since they would then form a trivial bundle.
\end{itemize}

Let us recap the trick we have used here to find an instanton with
nontrivial $\tilde w_2$ which manages to keep the entire $\spnh$
unbroken. Take, say, one exceptional $S^2$, call it $C_i$, and set
$\tilde w_2=C_i$. This value of $\tilde w_2$ is trivial is far as
$C_i$ is concerned since the self-intersection of $C_i$ is even. There
will be another curve $C_i^*$ dual to $C_i$, for which $C_i^*.C_i=
C_i^*.\tilde w_2=1$ and so $\tilde w_2$ is nontrivial over this
curve. Now go to the limit where we blow down $C_i$. We may put the
support of the curvature near $C_i$ as shown in \cite{BLPSSW:so32} and
so the curvature becomes zero everywhere except inside the point-like
instanton. Thus, all that matters for global holonomy are the
non-contractable loops around the lens space surrounding $C_i$. As we
have shown, this is trivial. It is important to notice that
$C_i^*$ is {\em not\/} blown down during this process and so does not
build a lens space which would pick up global holonomy.

In \cite{BLPSSW:so32} it was shown that a single obstructing instanton
locally contributes one to the second Chern class of the bundle. To
build our hidden obstructer we essentially double the value of
$F_{\mathbf{32}}$ in (\ref{eq:s32}). Since $c_2$ goes as the square of
the curvature, we see that we multiply the second Chern class by
four. That is, {\em the hidden obstructer contributes four to the
second Chern class}.

We thus know the two possibilities for producing compactifications of
the $\spnh$ heterotic string which preserve the $\spnh$ gauge
symmetry:
\begin{enumerate}
\item Point-like instantons at smooth points in the K3 surface which have
instanton number (i.e., contribution to $c_2$) one and $\tilde w_2=0$.
\item Point-like instantons stuck at orbifold points in the K3 surface
which have instanton number four and $\tilde w_2\neq0$.
\end{enumerate}

%%%%%%%%%%%%%%%%%%%%%%%%%%%%%%%%%%%%%%%%%%%%%%%%%%%%%%%%%%%%%%%%%%%

\section{The Dual Picture}    \label{s:F}

To understand nonperturbatively how string theory behaves on our new
instantons we need to find a dual picture. This is provided by
F-theory. Recall that F-theory associates a \CY\ threefold, $X$, to a
heterotic string on a K3 surface \cite{Vafa:F,MV:F}. For F-theory to
work, $X$ must be in the form of an elliptic fibration
$p:X\to\Theta$. One may regard the heterotic string on the K3 surface
as dual to either the type IIB string on $\Theta$ with some D-brane
insertions or, alternatively, to some special large radius limit of
the type IIA string on $X$. Either way, the non-perturbative physics
of the heterotic string becomes encoded in the elliptic fibration,
$p:X\to\Theta$. See \cite{MV:F2,me:lK3} for more details. In
particular, we will use the notation from \cite{me:lK3} and assume a
knowledge of many of the results in section 6 of that paper.

We would like to completely classify all heterotic string
compactifications on a K3 surface which lead to a gauge symmetry
containing $\spnh$ in the perturbatively-understood part of the gauge
symmetry. The only assumption we will make (subject to a few 
caveats outlined in \cite{me:lK3}) is that none of this gauge
symmetry arises from the Mordell-Weil group.

\def\HS#1{{\bf F}_{#1}}
\def\Ist#1{{\mathrm{I}\vphantom{\mathrm{II}}}^*_{#1}}
We wish to understand theories which at least begin as a
perturbatively-understood heterotic string theory. We thus want to
begin with one tensor multiplet and, as such, we assume $\Theta$ is of
the form of a Hirzebruch surface $\HS n$. 
The Hirzebruch surface is a $\P^1$-bundle over $\P_1$ with a natural
zero section, $C_0$. We will denote the class of the fibre, $f$. We
will call such fibres, ``$f$-curves'', to avoid any confusion with the
fibres of $X$ as an elliptic fibration.
To obtain an $\so(32)$ term
in the gauge algebra we require a line of $\Ist{12}$ fibres in
$\Theta$. To make this $\so(32)$ part of the perturbatively-understood
symmetry we put it along a section of $\HS n$. Let us assume it is the
zero section, $C_0$. We may do this without loss of generality so long
as we do {\em not\/} impose $n\geq0$.

To make the group precisely $\spnh$, it was shown in \cite{AG:sp32}
that one required $X$ to have precisely two global sections, as an
elliptic fibration. This forces a factorization of the Weierstrass
form of the elliptic fibration:
\begin{equation}
\begin{split}
y^2&=x^3+ax+b\\
a &= q - p^2\\
b &= -pq\\
\delta &= (q+2p^2)^2(4q-p^2),
\end{split}
\end{equation}
where $p$ and $q$ are functions over $\HS n$ and $\delta$ is the
discriminant. The $\Ist{12}$ condition forces $(a,b,\delta)$ to vanish
to order $(2,3,18)$ along $C_0$. Denote 
\begin{equation}
\begin{split}
m_1 &= q+2p^2\\
m_2 &= 4q-p^2.
\end{split}  \label{eq:m1m2}
\end{equation}
One may then show that $m_1$ must vanish to order 8 along $C_0$ and
$m_2$ vanishes to order 2. Let us use upper case letters to denote the
divisors in $\HS n$ associated to the various functions above. The \CY\
condition imposes
\begin{equation}
\begin{split}
\Delta &= 2M_1+M_2\\
M_1 = M_2 &= 8C_0 + (8+4n)f.
\end{split}
\end{equation}
Let us split off from $M_1$ and $M_2$ the parts giving the $\Ist{12}$
along $C_0$:
\begin{equation}
\begin{split}
M_1 &= M_1' + 8C_0\\
M_2 &= M_2' + 2C_0,
\end{split}
\end{equation}
with $\Delta'$ defined similarly. Now, to make sure that the fibres
along $C_0$ are generically nothing worse than $\Ist{12}$, neither
$M_1'$ nor $M_2'$ should contain any more of $C_0$. This means that
the intersection numbers
\begin{equation}
\begin{split}
M_1'.C_0 &= 8+4n\\
M_2'.C_0 &= 2(4-n),
\end{split}
\end{equation}
must be nonnegative. Thus $-2\leq n\leq 4$.

Let us treat the remainder of the discriminant, given by $M_1'$ and
$M_2'$ in turn. $M_1'$ is simply $8+4n$ copies of $f$. Generically
this will mean it is $8+4n$ parallel lines along the $f$ direction. As
$\Delta$ contains $2M_1'$, this will produce lines of $\mathrm{I}_2$
fibres. Thus, the gauge symmetry is enhanced nonperturbatively by $8+4n$
$\sp(1)$ terms. These are precisely Witten's simple point-like instantons of
\cite{W:small-i}.

If $k$ of these instantons are brought together, $k$ lines of
$\mathrm{I}_2$ will merge to form a line of $\mathrm{I}_{2k}$. As
explained in \cite{AG:sp32}, monodromy turns the $\su(2k)$ gauge
algebra one might first associate to this into an $\sp(k)$ gauge
algebra. One can potentially have monodromy whenever a curve in the
discriminant, whose associated gauge algebra may admit nontrivial
outer automorphisms, collides with another component of the
discriminant. Whether or not there is monodromy can be determined
purely in terms of the local geometry of the collision, and with what
type of curve it collided. In our case we have a transverse collision
of a line of $\Ist{12}$ fibres with a line of $\mathrm{I}_{2k}$
fibres. One may show that such a collision induces no monodromy in the
$\Ist{12}$ fibre but has a $\Z_2$ action in the $\mathrm{I}_{2k}$
fibre. We show how to determine how the monodromy acts in the
appendix.

As well as the gauge algebra, we may also determine the spectrum of
hypermultiplets as discussed in \cite{AG:sp32,BKV:enhg,KV:hyp}. The
transverse collision of the $\Ist{12}$ and $\mathrm{I}_{2k}$ produce a
half hypermultiplet in the $(\mbf{32},\mbf{2k})$ representation
of $\so(32)\oplus\sp(k)$. 

As discussed in \cite{W:small-i} we should
also expect a hypermultiplet in the $\mbf{k(2k-1)-1}$ (i.e.,
antisymmetric tensor) representation of $\sp(k)$. Call this the
$\mbf{A_2}$ representation for brevity. Let us use $\Delta''$ to
denote the discriminant after the contribution from $C_0$ {\em and\/}
all the $f$-curves has been subtracted.
To see how the hypermultiplets arise note that an $f$-curve is
topologically a sphere. Thus, if we are to have a 
nontrivial action of monodromy on the fibre of the elliptic fibration
around this sphere, we must have more than one branch point. At
present we have only found one collision --- that of $f$ with $C_0$. There must
be further collisions of $\Delta''$ with the $\mathrm{I}_{2k}$ line to
produce more monodromy. As explained by Morrison \cite{Mor:TASIF},
these collisions will produce the $\mbf{A_2}$ representation required.

To see this we use the results of \cite{KMP:enhg} which say that if
a curve of bad fibres is of genus $g$, then we expect $g$
hypermultiplets in the adjoint of the associated gauge algebra, in
addition to the usual adjoint of vectors. When monodromy acts within
the curve, the algebra is split between the part invariant under the
monodromy and the rest which varies. The vectors are only associated
with the monodromy-invariant part (see, for example, \cite{me:lK3})
but we may pick up hypermultiplets in the part that varies depending
on the genus of the base curve after we have taken the monodromy into
account. Thus, suppose we have a $\Z_2$ monodromy acting on a rational
curve in $\Theta$ associated to a gauge algebra (before monodromy is
taken into account) $\mathfrak{g}$. The outer automorphism induced
by the monodromy leaves $\mathfrak{g}_0$ invariant. The adjoint of
$\mathfrak{g}$ may then be decomposed into the adjoint of
$\mathfrak{g}_0$ plus a representation $R'$. Suppose the monodromy is
branched over $n_p$ points within the rational curve. Then as far as
the representation $R'$ is concerned, the base curve is actually a
double cover of the rational curve branched at $n_p$ points. This has
genus $\ff12n_p-1$. We therefore expect $\ff12n_p-1$ hypermultiplets
in the $R'$ representation.

In our case, we are reducing $\su(2k)$ to $\sp(k)$. It is easy to show
that $R'$ is indeed the $\mbf{A_2}$ representation. Now
we need to know how many points there are within each $f$-curve over
which the $\Z_2$ monodromy is branched. This will allow us to count
the $\mbf{A_2}$'s.

Let us introduce affine coordinates $(s,t)$ to parameterize $\HS n$
locally. Let $C_0$ be given by $s=0$ and let us fix a particular
$f$-curve to be given by $t=0$. To associate an $\sp(k)$ gauge
symmetry with this $f$-curve we require $m_1$ to be of order $k$ in
$t$. Let us put $m_1=t^k$ for the simplest case. Then
\begin{equation}
  \delta = t^{2k}(4t^k-9p^2).  \label{eq:simplek}
\end{equation}
Thus, $\Delta$ will collide with this $f$-curve whenever $p(s,t)$ has
a zero. $P$ is in the class $4C_0+(4+2n)f$ and thus collides with $f$
a total of $P.f=4$ times. One of these collisions is the transverse
collision with the line of $\Ist{12}$ fibres along $C_0$. The
other three collision are generically non-transverse collisions with a curve 
of $\mathrm{I}_1$ fibres along $\Delta''$. As we will see in the
appendix, all four collisions induce monodromy --- we have $n_p=4$
and thus one hypermultiplet in the $\mbf{A_2}$ representation as desired.

Thus far we have recovered the simple $\sp(1)$ point-like instantons.
Now we discover something new when we look at
collisions between $\Delta''$ and $C_0$. Since $M_1$ is order 8 along
$C_0$ as explained above, let us put $m_1=s^8$.
It follows that
\begin{equation}
  \delta = s^{16}(4s^8-9p^2).
\end{equation}
We know that $m_2$ vanishes to order 2 along $s=0$ so we must be able
to factorize $p=sp_1$. Therefore
\begin{equation}
  \delta = s^{18}(4s^6-9p_1^2).    \label{eq:p1C}
\end{equation}
Thus there will be collisions between $\Delta''$ and $C_0$ whenever
$p_1$ has extra zeros. This happens at $C_0.P_1=4-n$ points. 

Put $p_1=t$ to get a local form of the collision. (Note that now $t=0$ is
{\em not\/} the equation of an $f$-curve for simple instantons.)
Adding the degrees in $s$ and $t$ together, we see that $(a,b,\delta)$
have degrees $(4,6,20)$ respectively. As explained in \cite{me:lK3},
whenever these degrees are greater than, or equal to, $(4,6,12)$, one
must blow-up the base to resolve $X$. Therefore, these collision of
$\Delta''$ with $C_0$ induce new massless {\em tensor\/} degrees of
freedom.

Let $E_1$ be the resulting exceptional $\P^1$ in the blown-up
$\Theta$. The order to which $(a,b,\delta)$ vanish over generic points
in $E_1$ is given by subtracting $(4,6,12)$ from the orders at the
point which was blown-up. That is, the orders are $(0,0,8)$. Thus we
have $\mathrm{I}_8$ fibres along $E_1$. The collision between $E_1$
and $C_0$ produces monodromy. This results in a gauge symmetry of
$\sp(4)$.

\iffigs
\begin{figure}
$$
\setlength{\unitlength}{0.008750in}%
\begin{picture}(595,184)(40,610)
\thinlines
\put(385,660){\line( 1, 0){240}}
\put(485,620){\line( 0, 1){140}}
\multiput(465,740)(7.74194,0.00000){16}{\line( 1, 0){  3.871}}
\put( 60,660){\line( 1, 0){240}}
\multiput(180,620)(0.00000,7.80488){21}{\line( 0, 1){  3.902}}
\put(290,730){\vector( 1, 0){ 90}}
\put(500,715){
}
\put(500,715){
}
\put(220,670){
}
\put(205,730){\makebox(0,0)[lb]{$\Delta''$}}
\put(175,785){\makebox(0,0)[lb]{$f$}}
\put(450,730){\makebox(0,0)[lb]{$\tilde f$}}
\put(480,765){\makebox(0,0)[lb]{$E_1$}}
\put( 35,655){\makebox(0,0)[lb]{$C_0$}}
\put(285,637){\makebox(0,0)[lb]{$\Ist{12}$}}
\put(490,610){\makebox(0,0)[lb]{$\mathrm{I}_8$}}
\put(315,735){\makebox(0,0)[lb]{\scriptsize Blow up}}
\put(635,655){\makebox(0,0)[lb]{$C_0$}}
\end{picture}
$$
  \caption{Blowing up the $\Delta''$ collision with $C_0$.}
  \label{fig:bup1}
\end{figure}
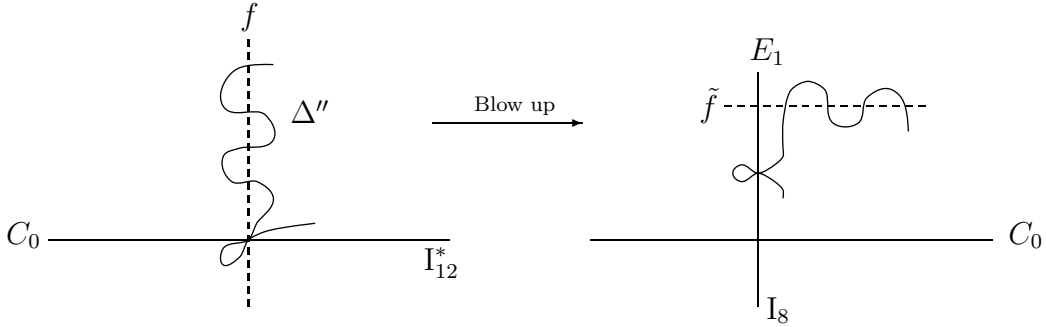
\fi

As shown in figure \ref{fig:bup1}, $\Delta''$ collides with $E_1$ just
once to produce another monodromy branch point. This means that
$n_p=2$ for this $\sp(4)$ gauge symmetry and so no hypermultiplets in
the $\mbf{A_2}$ representation appear. There will be hypermultiplets
in the $(\mbf{32},\mbf{8})$ representation from the collision of $E_1$ with
$C_0$. 

Let us review the spectrum we have obtained. Let the simple instantons
clump together in $\mu$ groups of $k_i$ (so that $\sum_{i=1}^{\mu}
k_i=8+4n$) but leave the collisions of $C_0$
and $\Delta''$ isolated. We have
\begin{enumerate}
\item A gauge algebra
\begin{equation}
  \so(32)\oplus\left(\bigoplus_{i=1}^\mu \sp(k_i)\right)
     \oplus\sp(4)^{\oplus(4-n)}.
\end{equation}
\item $5-n$ massless tensor supermultiplets (including the dilaton).
\item Hypermultiplets (or half-hypermultiplets if the representation
is not complex) in the following representations:
\begin{equation}
\begin{split}
  (\mbf{32},\mbf{2k_i})\quad&\mbox{of}\quad \so(32)\oplus \sp(k_i)\\
  \mbf{k_i(2k_i-1)-1}\quad&\mbox{of}\quad \sp(k_i)\\
  (\mbf{32},\mbf{8})\quad&\mbox{of}\quad \so(32)\oplus \sp(4) 
    \quad\quad\mbox{($4-n$ times)}
\end{split}
\end{equation}
as well as some chargeless hypermultiplets.
\end{enumerate}

At this point the interpretation of this model in terms of point-like
instantons discussed at the end of section \ref{s:ghol} should be
fairly evident. As mentioned above, the simple 
instantons are associated to the $8+4n$ zeros of $m_1$. Each of these
have instanton number one. If we assign instanton number four to each
of the $4-n$ collisions between $C_0$ and $\Delta''$ then
\begin{equation}
  \sum_{i=1}^{\mu} k_i + 4(4-n) = 24,
\end{equation}
for the total instanton number as expected for the bundle, $E$, on a
K3 surface. {\em We therefore identify these $4-n$ collisions as
point-like hidden-obstructer instantons in the dual heterotic string.}

We see that each hidden obstructer instanton is associated to a massless
tensor multiplet and an $\sp(4)$ gauge symmetry.

Let us be precise about what we mean exactly by this statement. As
discussed in \cite{SW:6d}, phase transitions between tensor moduli and
hypermultiplet moduli are fairly exotic in nature. This is exactly
what we have here --- when the size of the new instanton is shrunk
down to zero size (by hypermultiplets) a new modulus appears as the
scalar component of a tensor supermultiplet which allows us to move
off into a new component of the moduli space. Rather than speak of the
theory right at the phase transition point, which has ``tensionless
strings'' roughly speaking, we will assume that we switch on the new
tensor modulus slightly to move away from this peculiar theory. As a
result, we have a more conventional six-dimensional theory (although
it has no covariant action) and we may ask sensible questions about
anomalies etc. We shall not attempt to say anything in this paper
about the theory which sits right on the phase transition point.

To complete the spectrum we should count the number of chargeless
hypermultiplets. Roughly speaking, this is given by the number of
deformations of complex structure of $X$ plus one. One needs to be a little
careful however. It may be that F-theory counts some of the linear
combinations of charged hypermultiplets which can also act as
deformations. This latter effect is due to the appearance of
``elliptic scrolls'' in $X$ and is tied to Wilson's work on the
phenomenon of the K\"ahler cone jumping for special values of complex
structure \cite{Wil:Kc} (see also \cite{me:lK3} for a brief account
of this). This indeed happens when simple instantons coalesce. To
avoid this issue let us assume all $8+4n$ simple instantons
are isolated. 

We know $h^{1,1}(X)$ from the blow-ups in both the base (the tensor
multiplets) and the fibre (the rank of the gauge group). We have
$h^{1,1}=3+(4-n)+24+16=47-n$. To calculate $h^{2,1}(X)$ we need the
Euler characteristic of $X$. This is done by adding the contributions
from all of the bad fibres in $X$ as an elliptic fibration. For an
example see \cite{me:lK3}. In our case we need some Euler
characteristics of some of the fibres appearing over collisions within
$\Delta$. We calculate those required in the appendix. The result is%
\footnote{The only awkward step is calculating the Euler
characteristic of $\Delta''$. This is done by using the adjunction
formula and then compensating for the various high-order tacnodes
which appear in $\Delta''$.}
\begin{equation}
\begin{split}
\chi(X) &= \sum_{i=1}^{\mu}\left\{2k_i.(2-4)\right\}+18.(2-\mu-4+n)
+8.(4-n).(2-2) \\
&\qquad+ (-24-3n-3\mu-3\mu-(4-n))+3\sum_{i=1}^{\mu}(2+k_i)
+\sum_{i=1}^{\mu}(18+k_i)\\
&\qquad\qquad+(4-n).22+(4-n).6\\
&= 48-12n.
\end{split}
\end{equation}
This gives $h^{2,1}=h^{1,1}(X)-\ff12\chi(X)=23+5n$. Therefore there
are $24+5n$ chargeless hypermultiplets.

As always, one may check this F-theory calculation to ensure that
anomalies cancel (as they must). The gravitational anomaly yields
\begin{equation}
\begin{split}
273-29n_T-n_H+n_V&=273-29.(5-n)-\ff12.2.32.24-(24+5n)\\
  &\qquad+3.(8+4n)+36.(4-n)+496\\
  &=0.
\end{split}
\end{equation}
Similarly one may check the gauge anomalies.\footnote{I thank
N.~Seiberg for explaining this to me and showing that they do indeed
cancel.} 

This counting of chargeless hypermultiplets fits nicely with the
heterotic interpretation. The underlying K3 surface has 20
(quaternionic) deformations and the simple instantons may be placed
anywhere giving $8+4n$ more deformations. Each hidden obstructer
requires an orbifold point locally of the form $\C^2/\Z_2$, which
reduces the number of deformations of the K3 by one. The location of
each hidden obstructer is then fixed at this orbifold point. Thus, the
total number of deformations is $20+(8+4n)-(4-n)=24+5n$ as expected.

One may also check the above calculations in the case that some of the
simple instantons coalesce. In this case the topology of $X$ is
actually unchanged but the interpretation of some of the
hypermultiplets is modified.

%%%%%%%%%%%%%%%%%%%%%%%%%%%%%%%%%%%%%%%%%%%%%%%%%%%%%%%%%%%%%%%%%%%

\section{Some Equivalences}   \label{s:eq}

Recall the behaviour of the $E_8\times E_8$ heterotic string on a K3
surface as regards F-theory \cite{SW:6d,MV:F,MV:F2}. The topology
of the required $E_8\times E_8$-bundle on the K3 surface is specified
by how the total second Chern class is split between the two $E_8$'s. In
particular, F-theory on the Hirzebruch surface $\HS n$ is dual to a
split of $12+n$ and $12-n$.

This shows the T-duality between the $E_8\times E_8$ heterotic string
on a K3 surface and the $\spnh$ heterotic string on another K3
surface. For example, as has been known for some time \cite{MV:F}, the $\spnh$
heterotic string with $\tilde w_2=0$ must be dual to the $E_8\times
E_8$ string with the second Chern class split $(8,16)$ between the two
$E_8$'s. This follows since $\tilde w_2=0$ implies that there can be
no obstructers, hidden or non-hidden, which implies that $4-n=0$.

This raises a point which, at least at first sight, looks
puzzling. The global diffeomorphisms of the underlying K3 surface, on
which the heterotic string lives, can transform one value of $\tilde
w_2$ into another. In particular there are only three equivalence
classes once this is taken into account \cite{BLPSSW:so32}:
\begin{enumerate}
  \item $\tilde w_2=0$,
  \item $\tilde w_2\neq0$ and $\tilde w_2.\tilde w_2=0\pmod4$,
  \item $\tilde w_2\neq0$ and $\tilde w_2.\tilde w_2=2\pmod4$.
\end{enumerate}

Since $(4-n)$ hidden obstructers over disjoint $(-2)$-curves yields
$\tilde w_2.\tilde w_2=2(n-4)$ we are implying equivalences
between certain $E_8\times E_8$ string vacua. Actually these
equivalences do exist.

To see this we need to look at the strange properties of the
Hirzebruch surface.\footnote{I thank D.~Morrison for conversations
about this.} The topology of $\HS n$ is actually only specified by
whether $n$ is even or odd. Indeed one may build a family of surfaces
$\pi:Z\to D$, where $D$ is a complex disc with coordinate $z$ such
that the fibre at $z\neq0$ is $\HS n$ but at $z=0$ it becomes
$\HS{n+2}$. At $z=0$ a new algebraic curve within the fibre jumps into
existence with self-intersection $-n-2$. This causes the K\"ahler cone
to contract, relative to that of $\HS n$, but nothing has changed
topologically.

This equivalence between Hirzebruch surfaces is used to show the
equivalence of the $n=0$ model and the $n=2$ model as in
\cite{AG:mulK3,MV:F}. The elliptic threefold fibred over $\HS2$ is a
codimension one subset (which can be realized as a hypersurface in a
weighted projective space) of the more general member of the family
which is fibred over $\HS0$. The jumping K\"ahler cone of the
Hirzebruch surface is transfered to the threefold whose K\"ahler cone
also shrinks over this special sub-family.

It was shown in \cite{Wil:Kc} that jumping K\"ahler cones could only
happen in smooth threefolds if the algebraic class that jumped into
existence for special values of the complex structure was an
``elliptic scroll''. That is, an elliptic curve times a rational
curve. Thus, this rational curve is a $(0,-2)$-curve within the \CY\
threefold. Within the base of an elliptic fibration therefore, the
only curve which is allowed to jump into existence is a $(-2)$-curve,
which would come from the Hirzebruch surface $\HS2$. We
appear to have shown that the only equivalence allowed between models
is the $n=0$ to $n=2$ equivalence.

We may obtain the rest of the equivalences by relaxing the constraint
that the elliptic threefold be smooth. Now any smooth transition
between $\HS n$ and $\HS{n+2}$ may be turned into a ``smooth''
transition between singular elliptic threefolds. In the case we are
studying the \CY\ threefold has a curve of $D_{16}$-type singularities
inducing the $\spnh$ gauge symmetry. It is certainly singular.

Given this equivalence between \CY\ threefolds, there is no
contradiction between $\tilde w_2$ equivalence classes and F-theory
equivalence classes.

Now let us turn our attention to the connection between the hidden
obstructer theories we have described into terms of F-theory and other
models in the same $\tilde w_2$ class which break at least part of
$\spnh$. We focus on the Gimon-Polchinski models of \cite{GP:open}. As
explained in \cite{BLPSSW:so32}, we expect these models to all be in
the F-theory class with $n=0$.

To see this simply deform $m_2$ of (\ref{eq:m1m2}) so that it no
longer vanishes along $C_0$. This will turn the line of $\Ist{12}$
line of fibres along $C_0$ into a line of $\mathrm{I}_{16}$
fibres. This changes the class of $\Delta''$ but it will still collide
with $C_0$ at four points (doubly at each point). These collisions
will induce monodromy and so the $C_0$ line now generates a $\sp(8)$
gauge symmetry.

This breaking of $\spnh$ may be seen by the maximal subgroup
\begin{equation}
  \frac{\Spin(32)}{\Z_2} \supset
\SO(3)\times\frac{\Sp(8)}{\Z_2}\times\Z_2.
\end{equation}
Giving the hidden obstructer nonzero size can turn it into a smooth
$\SO(3)$-bundle. (The group must be non-simply-connected since $\tilde
w_2\neq0$.) Thus the global holonomy breaks the primordial gauge
symmetry to $\Sp(8)/\Z_2$ consistent with what we saw from F-theory.

Further deformations can be used to bunch the four points of collision
between $\Delta''$ and $C_0$ into two coalesced pairs. This will
remove the monodromy and so the gauge symmetry $\sp(8)$ will turn into
$\su(16)$.\footnote{Actually there are good reasons to expect the
Mordell-Weil group to enhance this further to $\gu(16)$. This is
because $X$ can be written as a K3 fibration whose generic fibre can
be written as a double cover of a rational elliptic surface. The
rational elliptic surface thus obtained is known from the
classification of \cite{Pers:RES} to have a Mordell-Weil group of rank
one. I thank M.~Gross for conversations on this point.}
To fit in with the work of \cite{GP:open} (see also the earlier work
of \cite{BSag:u16}) we may then identify the two
points of collision of $\Delta''$ with $C_0$ as each yielding a
hypermultiplet in the $\mbf{120}$ of $\su(16)$.

The line of fibres along $C_0$ can be broken up into a parallel set of
lines of $\mathrm{I}_{2l_j}$ fibres so that $\sum_jl_j=8$. This makes
the class $C_0$ analogous to the class $f$ in which we have a set of
parallel lines of $\mathrm{I}_{2k_i}$ fibres satisfying
$\sum_ik_i=8$. This is as it should be since $\HS0$ has an obvious
symmetry between the classes $C_0$ and $f$.

This allows us to reproduce all of the Gimon-Polchinski models in
terms of F-theory. We show an example in figure \ref{fig:GP}. Note
that the collisions within the discriminant will produce massless
hypermultiplets in various representations in the usual way.

\iffigs
\begin{figure}
$$
\setlength{\unitlength}{0.008750in}%
\begin{picture}(370,355)(80,445)
\thinlines
\put(240,505){\line( 1, 0){ 10}}
\put(290,505){\line( 1, 0){ 10}}
\put(340,505){\line( 1, 0){ 10}}
\put(200,505){
}
\put(250,505){
}
\put(300,505){
}
\put(350,505){
}
\put( 80,625){\line( 0, 1){ 10}}
\put( 80,675){\line( 0, 1){ 10}}
\put( 80,725){\line( 0, 1){ 10}}
\put( 80,585){
}
\put( 80,635){
}
\put( 80,685){
}
\put( 80,735){
}
\put(100,800){\line( 0,-1){335}}
\put(160,800){\line( 0,-1){335}}
\put( 80,485){\line( 1, 0){360}}
\put( 80,565){\line( 1, 0){360}}
\put(210,595){
}
\put(310,595){
}
\put(190,740){
}
\put(190,670){
}
\put(190,720){
}
\put(230,595){
}
\put(450,480){\makebox(0,0)[lb]{$\mathrm{I}_{2l_1}$}}
\put(450,560){\makebox(0,0)[lb]{$\mathrm{I}_{2l_2}$}}
\put(155,445){\makebox(0,0)[lb]{$\mathrm{I}_{2k_2}$}}
\put( 95,445){\makebox(0,0)[lb]{$\mathrm{I}_{2k_1}$}}
\put(380,570){\makebox(0,0)[lb]{$\su(2l_2)$}}
\put(400,490){\makebox(0,0)[lb]{$\sp(l_1)$}}
\put(105,780){\makebox(0,0)[lb]{$\sp(k_1)$}}
\put(165,780){\makebox(0,0)[lb]{$\su(2k_2)$}}
\put(330,630){\makebox(0,0)[lb]{$\Delta''$}}
\end{picture}
$$
  \caption{F-theory picture of Gimon-Polchinski models.}
  \label{fig:GP}
\end{figure}
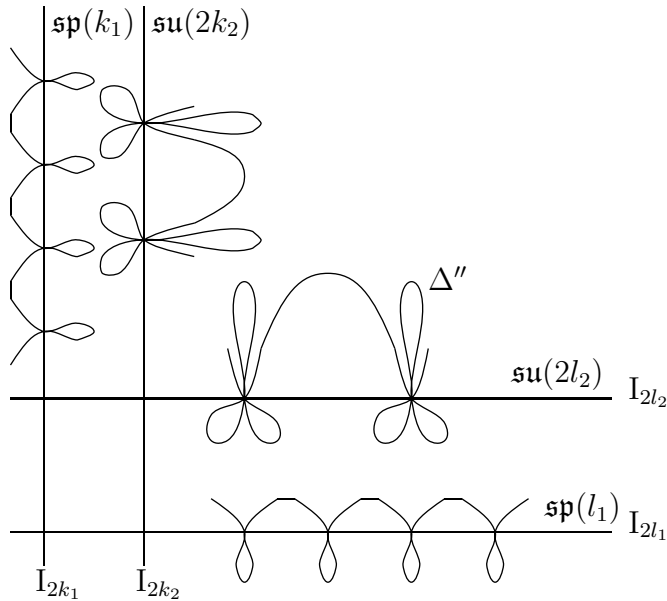
\fi

%%%%%%%%%%%%%%%%%%%%%%%%%%%%%%%%%%%%%%%%%%%%%%%%%%%%%%%%%%%%%%%%%%%

\section{Phase Transitions}   \label{s:ph}

Let us return to the $E_8\times E_8$ heterotic string with its second
Chern class split as $12+n$ and $12-n$ between the two $E_8$'s.
When any of the instantons become point-like in the $E_8\times E_8$ a
new massless tensor supermultiplet appears. One may then use this new
direction in the moduli space to move to another theory with a
point-like instanton with the second Chern class split
$(11+n,13-n)$. This instanton can then be given size to remove the
massless tensor. Thus, by a process that involves two phase
transitions, we may effectively change the topology of the $E_8\times
E_8$-bundle. 

In terms of M-theory \cite{SW:6d} this was understood by a 5-brane
peeling itself off one ``end of the universe'' and moving over (by
varying the tensor degree of freedom) to the other end of the
universe. 

In terms of F-theory \cite{MV:F2}, which is the approach we use here,
this is achieved by first blowing up a point in $\HS n$. The proper
transform of the fibre that passed through this point then has
self-intersection $-1$ allowing it to be blown down. This blow down
results in the Hirzebruch surface $\HS {n\pm 1}$ (depending on whether
the original point blown up was on $C_0$ or not).

Our new point-like hidden obstructer instanton is very similar is the
point-like $E_8$ instanton in that a new massless tensor results. We
may therefore follow a phase transition to another Hirzebruch surface
and see what happens. We will discover that we may transform hidden
obstructer instantons into simple instantons and {\em vice versa}.

Begin with the collision of $\Delta''$ with $C_0$ in the Hirzebruch
surface, $\HS n$, as in the previous section. As we discussed above, to
resolve $X$, such a collision must be blown-up within $\HS n$. The
exceptional divisor, $E_1$, results. Let $\tilde f$ be the proper transform
of the $f$-curve that passed through the collision. As blow-ups
decrease self-intersections by one and $f.f=0$, we see that
$\tilde f.\tilde f=-1$. 

We know that $\Delta''.f=6$ and that the collision with $C_0$ accounts
for two of these intersections. Thus, away from $C_0$, $\Delta''$ hits
our particular $f$-curve four times. Assuming everything else is
generic, these will be at four distinct points. Thus the proper transform
of $\Delta''$, which we also denote $\Delta''$, hits $\tilde f$ at
four distinct points.

We may now blow down $\tilde f$. This gives the proper transform of
$E_1$ a self-intersection of 0 and it becomes a fibre, $f$, of the
Hirzebruch surface $\HS{n+1}$ which we have now made. Now $\Delta''$
will hit this new $f$-curve four times at the same point (where
$\tilde f$ used to hit $E_1$). 

Now deform $X$ so that this quadruple collision of $\Delta''$ with $f$
divides into two double collisions. What we have done is to produce
exactly the F-theory picture of four coalesed simple instantons giving
a gauge group $\sp(4)$.

Let us repeat what we have done in the language of the heterotic
string. Begin with a point-like hidden obstructer instanton. Then
move along in moduli space from one phase to another using the
massless tensor degree of freedom. Then deform using hypermultiplets
to get rid of the massless tensor. The hidden obstructer has
disappeared ($n$ has increased by one) but four new simple
instantons have appeared.

We see therefore that our two types of point-like instantons may be
transformed into each other by using massless tensors. This also gives
a way of changing the topology (as given by $\tilde w_2$) of the
associated vector bundle. Thus we see that the picture is very
analogous to the $E_8\times E_8$ heterotic string.

Note that the geometry of the K3 surface is given by hypermultiplet
deformations and so is fixed while we vary the tensor. As we knew that
the hidden obstructer lived on an orbifold point, the orbifold point
must still be there after moving along the tensor direction. What's
more we know that the location of the simple instantons are also given
by hypermultiplets. This means that, before we get rid of the massless
tensor by moving the simple instantons, the four simple instantons
must have been sat right on the orbifold point. This implies that {\em
four simple instantons on a $\C^2/\Z_2$ quotient singularity in the K3
surface produce a massless tensor supermultiplet.}

\iffigs
\begin{figure}
$$
\setlength{\unitlength}{0.008750in}%
\begin{picture}(566,621)(24,200)
\thinlines
\put( 80,800){\line( 0,-1){160}}
\put( 60,780){
}
\put(130,800){\line( 0,-1){160}}
\put(110,780){
}
\put(180,800){\line( 0,-1){160}}
\put(160,780){
}
\put(230,800){\line( 0,-1){160}}
\put(210,780){
}
\put(280,800){\line( 0,-1){160}}
\put(260,780){
}
\put(420,800){\line( 0,-1){160}}
\put(400,780){
}
\put( 80,580){\line( 0,-1){160}}
\put( 60,560){
}
\put(220,360){\line( 0,-1){160}}
\put(200,340){
}
\put(200,240){\line( 1, 0){240}}
\multiput(320,200)(0.00000,7.80488){21}{\line( 0, 1){  3.902}}
\put(360,250){
}
\put( 40,680){\line( 1, 0){270}}
\put(380,680){\line( 1, 0){200}}
\put(520,800){\line( 0,-1){160}}
\put(310,740){\vector( 1, 0){ 75}}
\put(470,640){\vector( 0,-1){ 60}}
\put(380,530){\vector(-1, 0){100}}
\put(380,460){\line( 1, 0){200}}
\put(520,580){\line( 0,-1){160}}
\put( 60,460){\line( 1, 0){240}}
\put(160,420){\line( 0, 1){140}}
\multiput(140,540)(7.74194,0.00000){16}{\line( 1, 0){  3.871}}
\put(420,580){\line( 0,-1){160}}
\put(135,795){\vector( 1, 0){ 30}}
\put(220,795){\vector(-1, 0){ 25}}
\put(270,795){\vector(-1, 0){ 25}}
\put(200,430){\vector( 3,-4){ 45}}
\put(505,785){
}
\put(505,760){
}
\put(505,730){
}
\put(175,515){
}
\put(175,515){
}
\put(400,560){
}
\put(492,566){
}
\put( 14,675){\makebox(0,0)[lb]{$C_0$}}
\put(299,658){\makebox(0,0)[lb]{$\Ist{12}$}}
\put(585,675){\makebox(0,0)[lb]{$C_0$}}
\put( 32,775){\makebox(0,0)[lb]{$\Delta''$}}
\put(540,739){\makebox(0,0)[lb]{$\Delta''$}}
\put( 75,620){\makebox(0,0)[lb]{$\mathrm{I}_2$}}
\put(125,620){\makebox(0,0)[lb]{$\mathrm{I}_2$}}
\put(175,620){\makebox(0,0)[lb]{$\mathrm{I}_2$}}
\put(225,620){\makebox(0,0)[lb]{$\mathrm{I}_2$}}
\put(275,620){\makebox(0,0)[lb]{$\mathrm{I}_2$}}
\put(325,745){\makebox(0,0)[lb]{\scriptsize Coalesce}}
\put(480,600){\makebox(0,0)[lb]{\scriptsize Make K3 orbifold}}
\put(310,535){\makebox(0,0)[lb]{\scriptsize Blow up}}
\put(230,400){\makebox(0,0)[lb]{\scriptsize Blow down $E_1$}}
\put(525,630){\makebox(0,0)[lb]{$\mathrm{I}_8$}}
\put(345,310){\makebox(0,0)[lb]{$\Delta''$}}
\put(445,235){\makebox(0,0)[lb]{$C_0$}}
\put(125,535){\makebox(0,0)[lb]{$\tilde f$}}
\put(318,365){\makebox(0,0)[lb]{$f$}}
\put(517,807){\makebox(0,0)[lb]{$E_1$}}
\put(155,565){\makebox(0,0)[lb]{$E_1$}}
\end{picture}
$$
  \caption{Four simple instantons collide with an orbifold point.}
  \label{fig:phase}
\end{figure}
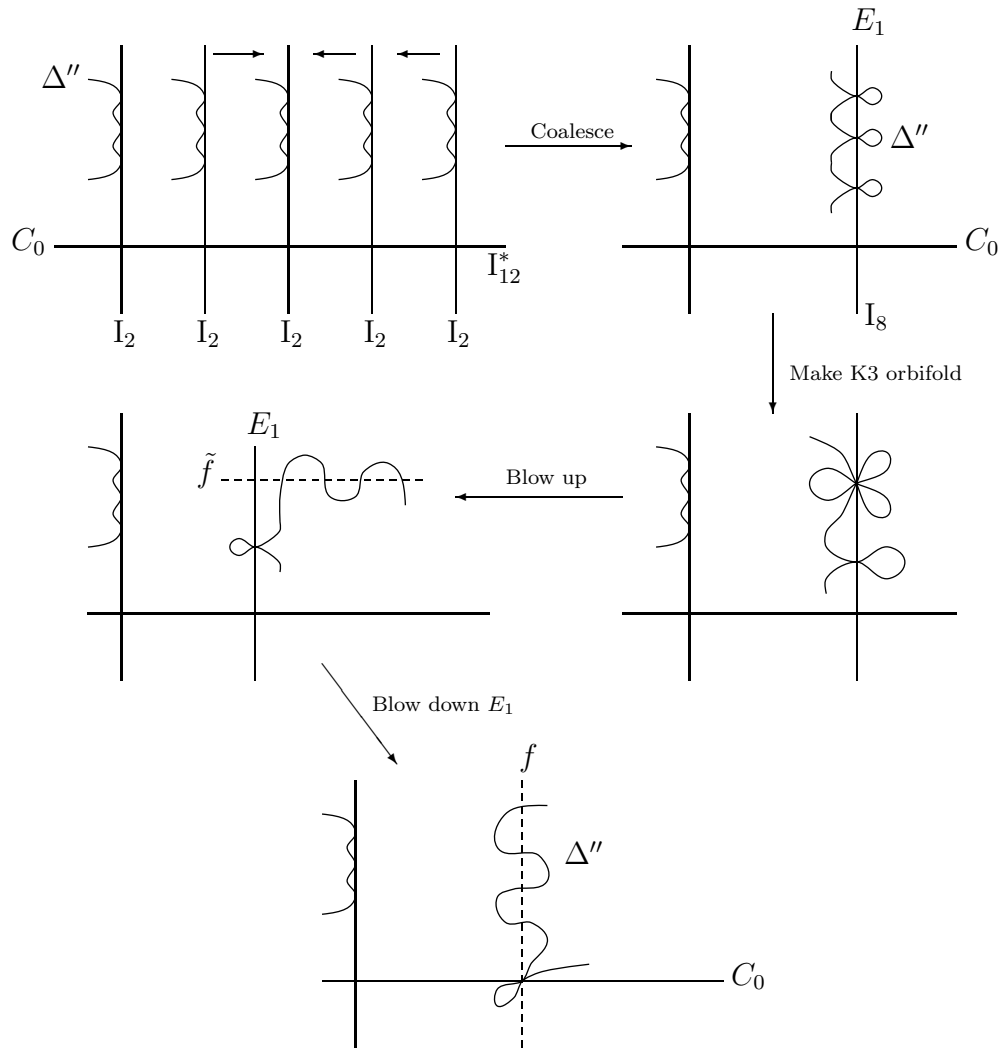
\fi

In figure \ref{fig:phase} we show the phase transition we described
above backwards. Start with a heterotic string with, say $\tilde
w_2=0$ (and therefore $\Theta\cong\HS4$). Then bring four simple
point-like instantons together to form 
a theory containing a gauge symmetry of $\sp(4)$. Now give the K3
surface a $\C^2/\Z_2$ quotient singularity and put this coalesced
instanton at that point. Now a massless tensor appears. Use this
massless tensor to turn $\HS 4$ into $\HS 3$. Now we have a hidden
obstructer instanton at the orbifold point. If we wish, the massless
tensor may be given mass by giving the new point-like instanton size
(which will break $\so(32)$).

A natural question to ask is what happens if fewer than four simple
instantons coalesce at an orbifold point. Let us consider $k$ simple
instantons. The collision of the associated $f$-curve with $\Delta''$
was given in (\ref{eq:simplek}). Consider the zeros of $p$ as $s$ is
varied to move along $f$. Generically the zeros are isolated. It is
evident from the above discussion that the orbifold condition amounts
to $p$ having a zero of order two. We are therefore interested in a
collision roughly of the form
\begin{equation}
\begin{split}
  a &= t^k-3s^4\\
  b &= -s^2(t^k-2s^4)\\
  \delta &= t^{2k}(4t^k-9s^4).
\end{split}
\end{equation}
Adding the degrees of $s$ and $t$ together we see that $(a,b,\delta)$
have degrees $(\min(k,4),2+\min(k,4),2k+\min(k,4))$ respectively. Thus
we hit the required $(4,6,12)$ for a massless tensor precisely when
$k\geq4$. That is, fewer than four simple instantons at an orbifold
point are not enough to produce the massless tensor. 

%%%%%%%%%%%%%%%%%%%%%%%%%%%%%%%%%%%%%%%%%%%%%%%%%%%%%%%%%%%%%%%%%%%

\section{Coalesced Instantons}    \label{s:co}

Now we know that four simple instantons at an orbifold point produce a
massless tensor which connects the theory to a hidden obstructer, the
natural question to ask is what happens when more than four simple
instantons coalesce at an orbifold point. This is equivalent to asking
what happens when a simple instanton hits a hidden obstructer. It is
then natural to ask what happens when two hidden obstructers coalesce.

\subsection{A simple instanton meets a hidden obstructer}
    \label{ss:sh}

Let $k$ simple instantons hit a hidden obstructer. Recall that a
hidden obstructer corresponds to a collision of $\Delta''$ with
$C_0$. Consider the $f$-curve passing through this collision
point. From our discussion of simple instantons above and their
relationship to $M_1$, it is clear that we require $M_1$ to contain
$k$ times this $f$-curve. That is, $\Delta$ includes $2k$ times this
$f$-curve.

Following (\ref{eq:p1C}), the form of the discriminant is
\begin{equation}
  \delta = s^{18}t^{2k}(4s^6-9p_1^2),
\end{equation}
where $p_1$, where $s=0$, has a single zero at $t=0$. Adding the
degrees of $s$ and $t$ together gives the degrees of $(a,b,\delta)$
equal to $(4,6,20+2k)$. Thus we have a blow-up in the base. Now the
degrees along the exceptional divisor, $E_1$, are $(0,0,8+2k)$. This gives a
gauge symmetry $\sp(4+k)$. The proper transform, $\tilde f$, of the
$f$-curve that 
passed through the collision is still a line of $\mathrm{I}_{2k}$
fibres and so we also have an $\sp(k)$ gauge symmetry. Since this curve
hits $E_1$, we expect a hypermultiplet in the $(\mbf{8+2k},\mbf{2k})$
representation of the
$\sp(4+k)\oplus\sp(k)$ part of the gauge algebra. 
This can be seen by applying monodromy to the results of \cite{BSV:D-man}.
We also have hypermultiplets from
the $C_0$ collision with $E_1$ but there is no collision between $C_0$
and $\tilde f$.

\iffigs
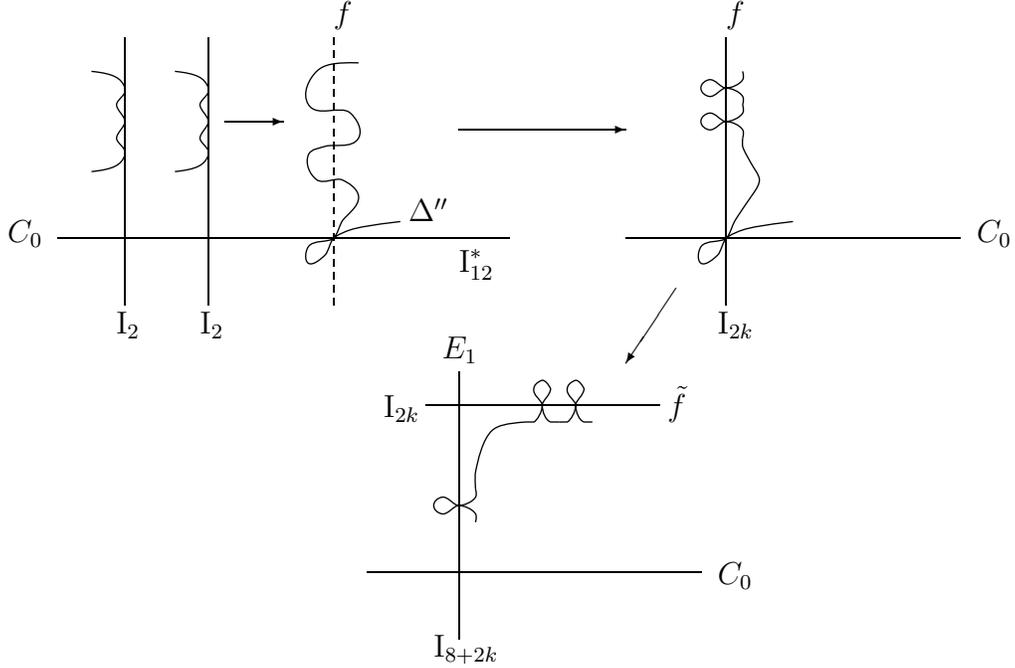
\begin{figure}
$$
\setlength{\unitlength}{0.008750in}%
\begin{picture}(570,394)(20,425)
\thinlines
\put( 80,800){\line( 0,-1){160}}
\put( 60,780){
}
\put(130,800){\line( 0,-1){160}}
\put(110,780){
}
\multiput(205,640)(0.00000,7.80488){21}{\line( 0, 1){  3.902}}
\put(245,690){
}
\put( 40,680){\line( 1, 0){270}}
\put(380,680){\line( 1, 0){200}}
\put(225,480){\line( 1, 0){200}}
\put(440,640){\line( 0, 1){160}}
\put(280,440){\line( 0, 1){160}}
\put(260,580){\line( 1, 0){140}}
\put(140,750){\vector( 1, 0){ 35}}
\put(280,745){\vector( 1, 0){100}}
\put(410,650){\vector(-2,-3){ 30}}
\put(480,690){
}
\put(450,780){
}
\put(450,760){
}
\put(450,740){
}
\put(290,530){
}
\put(340,570){
}
\put(360,570){
}
\put(290,530){
}
\put( 75,620){\makebox(0,0)[lb]{$\mathrm{I}_2$}}
\put(125,620){\makebox(0,0)[lb]{$\mathrm{I}_2$}}
\put(205,805){\makebox(0,0)[lb]{$f$}}
\put(250,690){\makebox(0,0)[lb]{$\Delta''$}}
\put( 10,675){\makebox(0,0)[lb]{$C_0$}}
\put(280,655){\makebox(0,0)[lb]{$\Ist{12}$}}
\put(590,675){\makebox(0,0)[lb]{$C_0$}}
\put(440,805){\makebox(0,0)[lb]{$f$}}
\put(405,570){\makebox(0,0)[lb]{$\tilde f$}}
\put(435,620){\makebox(0,0)[lb]{$\mathrm{I}_{2k}$}}
\put(235,570){\makebox(0,0)[lb]{$\mathrm{I}_{2k}$}}
\put(265,425){\makebox(0,0)[lb]{$\mathrm{I}_{8+2k}$}}
\put(270,605){\makebox(0,0)[lb]{$E_1$}}
\put(435,470){\makebox(0,0)[lb]{$C_0$}}
\end{picture}
$$
  \caption{$k$ simple instantons hitting a hidden obstructer.}
  \label{fig:sh}
\end{figure}
\fi

As shown in figure \ref{fig:sh} and the appendix, there are only two points of
monodromy in the curves generating both the $\sp(4+k)$ and the
$\sp(k)$ gauge algebras. Thus we have no hypermultiplets in the
$\mbf{A_2}$ representations of either of these algebras.

As an example, suppose $k_1$ of the simple point-like instantons
collide with one of the $4-n$ hidden obstructers and let the remaining
$8+4n-k_1$ clump into groups of $k_i$, $i=2\ldots\mu$. The spectrum is
\begin{enumerate}
\item A gauge algebra
\begin{equation}
  \so(32)\oplus\sp(k_1)\oplus\sp(4+k_1)
  \oplus\left(\bigoplus_{i=2}^\mu \sp(k_i)\right)
     \oplus\sp(4)^{\oplus(3-n)}.
\end{equation}
\item $5-n$ massless tensor supermultiplets (including the dilaton).
\item Hypermultiplets (or half-hypermultiplets if the representation
is not complex) in the following representations:
\begin{equation}
\begin{split}
  (\mbf{32},\mbf{8+2k_1})\quad&\mbox{of}\quad \so(32)\oplus\sp(4+k_1)\\
  (\mbf{8+2k_1},\mbf{2k_1})\quad&\mbox{of}\quad\sp(4+k_1)\oplus\sp(k_1)\\
  (\mbf{32},\mbf{2k_i})\quad&\mbox{of}\quad \so(32)\oplus \sp(k_i)\\
  \mbf{k_i(2k_i-1)-1}\quad&\mbox{of}\quad \sp(k_i)\\
  (\mbf{32},\mbf{8})\quad&\mbox{of}\quad \so(32)\oplus \sp(4) 
    \quad\quad\mbox{($3-n$ times)},
\end{split}
\end{equation}
where $i=2\ldots\mu$,
as well as $20+(\mu-1)-(4-n)$ chargeless hypermultiplets.
\end{enumerate}
The reader may check that anomalies cancel.

\subsection{Two hidden obstructers meet}   \label{ss:hh}

The natural thing to identify with two coalesced hidden obstructers is
when two of the zeroes of $p_1$ in (\ref{eq:p1C}) coalesce.
This may be achieved by putting $p_1=t^2+\alpha st+\beta s^2$ for some
generic $\alpha,\beta$. We obtain a total degree for $(a,b,\delta)$ at
$s=t=0$ equal to $(6,9,22)$ respectively. When we blow up this point,
we obtain the exceptional divisor, $E_1$, with degrees
$(2,3,10)$. Thus $E_1$ is a curve of $\Ist{4}$. There
is no monodromy within this and so a gauge algebra $\so(16)$ results.

We are not done however. The collision of the curve of $\Ist{12}$
fibres along $C_0$ and $\Ist{4}$ fibres along $E_1$ has total degree
$(4,6,28)$. Therefore we are required to blow-up this point too. This
introduces an exceptional divisor $E_2$. Along this curve, the degrees
are $(0,0,16)$. In this case there is monodromy and so the gauge
algebra is $\sp(8)$. Finally there are two collision of $E_1$ with the proper
transform of $\Delta''$ which also require blowing up. The collisions
each have total degree $(4,6,12)$ and so the resulting two exceptional
divisors, $E_3$ and $E_4$, carry smooth fibres and hence no
further gauge algebra. See figure \ref{fig:hh} for this process.

\iffigs
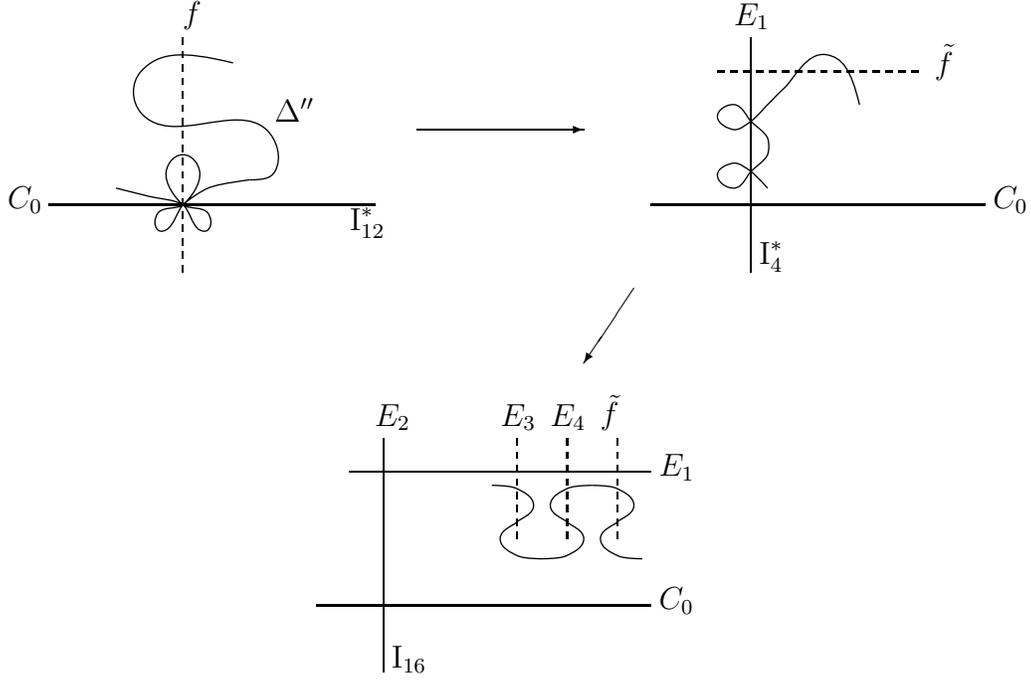
\begin{figure}
$$
\setlength{\unitlength}{0.008750in}%
\begin{picture}(580,399)(45,420)
\thinlines
\put(220,460){\line( 1, 0){200}}
\put(260,560){\line( 0,-1){140}}
\put(240,540){\line( 1, 0){180}}
\multiput(340,560)(0.00000,-8.00000){8}{\line( 0,-1){  4.000}}
\multiput(370,560)(0.00000,-8.00000){8}{\line( 0,-1){  4.000}}
\multiput(400,560)(0.00000,-8.00000){8}{\line( 0,-1){  4.000}}
\put(280,745){\vector( 1, 0){100}}
\put(410,650){\vector(-2,-3){ 30}}
\put( 60,700){\line( 1, 0){195}}
\multiput(140,800)(0.00000,-8.00000){18}{\line( 0,-1){  4.000}}
\put(420,700){\line( 1, 0){200}}
\put(480,800){\line( 0,-1){140}}
\multiput(460,780)(7.74194,0.00000){16}{\line( 1, 0){  3.871}}
\put(100,710){
}
\put(180,715){
}
\put(325,532){
}
\put(385,532){
}
\put(385,532){
}
\put(490,710){
}
\put(490,760){
}
\put( 35,695){\makebox(0,0)[lb]{$C_0$}}
\put(240,680){\makebox(0,0)[lb]{$\Ist{12}$}}
\put(195,750){\makebox(0,0)[lb]{$\Delta''$}}
\put(140,805){\makebox(0,0)[lb]{$f$}}
\put(590,775){\makebox(0,0)[lb]{$\tilde f$}}
\put(625,695){\makebox(0,0)[lb]{$C_0$}}
\put(470,805){\makebox(0,0)[lb]{$E_1$}}
\put(485,660){\makebox(0,0)[lb]{$\Ist{4}$}}
\put(425,455){\makebox(0,0)[lb]{$C_0$}}
\put(255,565){\makebox(0,0)[lb]{$E_2$}}
\put(265,420){\makebox(0,0)[lb]{$\mathrm{I}_{16}$}}
\put(425,535){\makebox(0,0)[lb]{$E_1$}}
\put(330,565){\makebox(0,0)[lb]{$E_3$}}
\put(360,565){\makebox(0,0)[lb]{$E_4$}}
\put(390,565){\makebox(0,0)[lb]{$\tilde f$}}
\end{picture}
$$
  \caption{Two hidden obstructers at the same point.}
  \label{fig:hh}
\end{figure}
\fi

It is not much harder to go directly to the case of $k$ more simple
instantons joining the two coalesced hidden obstructers. In this case
the $f$-curve passing through the complicated collision of $\Delta''$
with $C_0$ will now carry $\mathrm{I}_{2k}$. Now the blow-up process
is similar to the above case except that more singular fibres
appear. This process is shown in figure \ref{fig:hhs}.

\iffigs
\begin{figure}
$$
\setlength{\unitlength}{0.008750in}%
\begin{picture}(585,400)(45,420)
\thinlines
\multiput(180,800)(0.00000,-8.00000){18}{\line( 0,-1){  4.000}}
\put(140,710){
}
\put(220,715){
}
\put(280,745){\vector( 1, 0){100}}
\put( 60,700){\line( 1, 0){195}}
\put(435,700){\line( 1, 0){195}}
\put(420,460){\line( 1, 0){200}}
\put(480,560){\line( 0,-1){140}}
\put( 45,460){\line( 1, 0){200}}
\put( 85,560){\line( 0,-1){140}}
\put( 65,540){\line( 1, 0){180}}
\put( 80,820){\line( 0,-1){160}}
\put(120,820){\line( 0,-1){160}}
\put(125,795){\vector( 1, 0){ 35}}
\put(515,800){\line( 0,-1){140}}
\put(515,650){\vector( 0,-1){ 70}}
\put(380,520){\vector(-1, 0){100}}
\put(460,540){\line( 1, 0){120}}
\put(165,560){\line( 0,-1){ 60}}
\put(195,560){\line( 0,-1){ 60}}
\put(225,560){\line( 0,-1){ 60}}
\put(475,710){
}
\put( 60,810){
}
\put(100,810){
}
\put(555,715){
}
\put(500,470){
}
\put(150,530){
}
\put(180,530){
}
\put(210,530){
}
\put( 35,695){\makebox(0,0)[lb]{$C_0$}}
\put(240,680){\makebox(0,0)[lb]{$\Ist{12}$}}
\put(410,695){\makebox(0,0)[lb]{$C_0$}}
\put(615,680){\makebox(0,0)[lb]{$\Ist{12}$}}
\put(515,805){\makebox(0,0)[lb]{$f$}}
\put(235,750){\makebox(0,0)[lb]{$\Delta''$}}
\put(180,805){\makebox(0,0)[lb]{$f$}}
\put(580,535){\makebox(0,0)[lb]{$\tilde f$}}
\put(625,455){\makebox(0,0)[lb]{$C_0$}}
\put(475,565){\makebox(0,0)[lb]{$E_1$}}
\put(250,455){\makebox(0,0)[lb]{$C_0$}}
\put( 80,565){\makebox(0,0)[lb]{$E_2$}}
\put( 90,420){\makebox(0,0)[lb]{$\mathrm{I}_{16+2k}$}}
\put(250,535){\makebox(0,0)[lb]{$E_1$}}
\put(160,565){\makebox(0,0)[lb]{$E_3$}}
\put(190,565){\makebox(0,0)[lb]{$E_4$}}
\put(220,565){\makebox(0,0)[lb]{$\tilde f$}}
\put(520,660){\makebox(0,0)[lb]{$\mathrm{I}_{2k}$}}
\put( 80,640){\makebox(0,0)[lb]{$\mathrm{I}_{2}$}}
\put(120,640){\makebox(0,0)[lb]{$\mathrm{I}_{2}$}}
\put(485,420){\makebox(0,0)[lb]{$\Ist{4+2k}$}}
\put(550,545){\makebox(0,0)[lb]{$\mathrm{I}_{2k}$}}
\put(160,480){\makebox(0,0)[lb]{$\mathrm{I}_{2k}$}}
\put(190,480){\makebox(0,0)[lb]{$\mathrm{I}_{2k}$}}
\put(220,480){\makebox(0,0)[lb]{$\mathrm{I}_{2k}$}}
\end{picture}
$$
  \caption{Two hidden obstructers meet $k$ simple instantons.}
  \label{fig:hhs}
\end{figure}
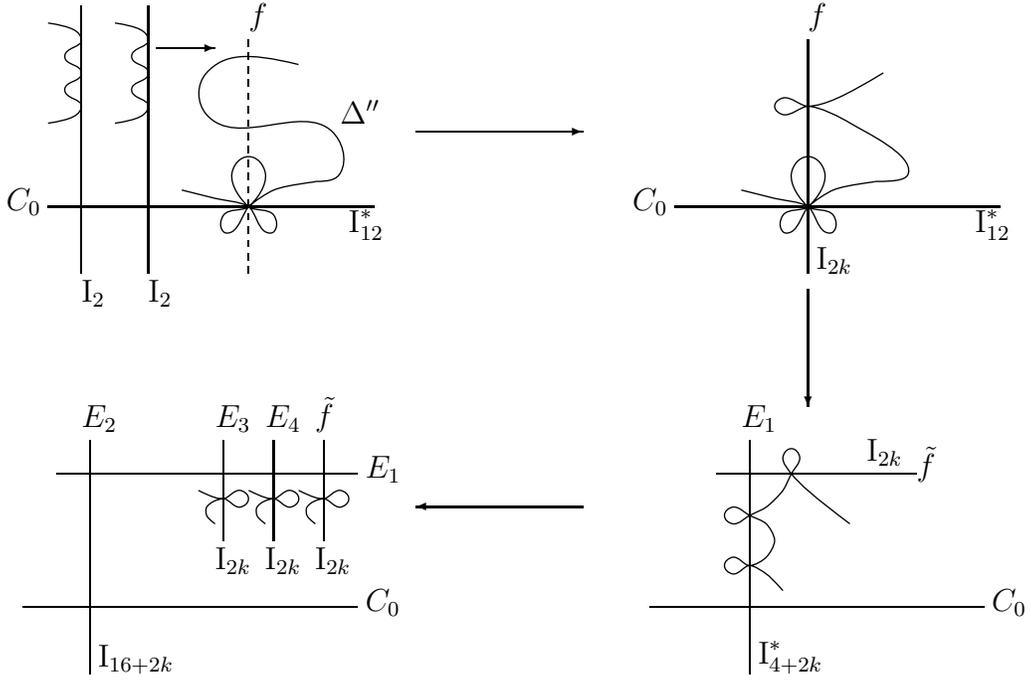
\fi

Let us give the spectrum that results in this case. Let the remaining
$8+4n-k$ simple instantons that have not joined the hidden obstructers
be disjoint. It is an easy matter to generalize to the case where
these coalesce amongst themselves but it will clutter the notation.
The result is
\begin{enumerate}
\item A gauge algebra
\begin{equation}
  \sp(1)^{\oplus(8+4n-k)}\oplus\so(32)\oplus\sp(8+k)\oplus
  \so(16+4k)\oplus\sp(k)^{\oplus3}\oplus\sp(4)^{\oplus(2-n)}.
\end{equation}
\item $7-n$ massless tensor supermultiplets (including the dilaton).
\item Hypermultiplets (or half-hypermultiplets if the representation
is not complex) in the following representations:
\begin{equation}
\begin{split}
  (\mbf{32},\mbf{2})\quad&\mbox{of}\quad \so(32)\oplus \sp(1)
    \quad\quad\mbox{($8+4n-k$ times)}\\
  (\mbf{32},\mbf{16+2k})\quad&\mbox{of}\quad \so(32)\oplus \sp(8+k)\\
  (\mbf{16+4k},\mbf{16+2k})\quad&\mbox{of}\quad \so(16+4k)\oplus \sp(8+k)\\
  (\mbf{16+4k},\mbf{2k})\quad&\mbox{of}\quad \so(16+4k)\oplus \sp(k)
    \quad\quad\mbox{($3$ times)}\\
  (\mbf{32},\mbf{8})\quad&\mbox{of}\quad \so(32)\oplus \sp(4) 
    \quad\quad\mbox{($2-n$ times)},
\end{split}
\end{equation}
as well as $22+5n-k$ chargeless hypermultiplets.
\end{enumerate}
As usual the anomalies miraculously cancel.

A couple of points are worth noting. Firstly the gauge group is
getting pretty large. For example, putting $n=2$ and $k=16$ in the
above yields a rank 128 gauge group. It also contains an $\so(80)$
factor in this case. This is interesting as we know that a rank 40
gauge symmetry can never be understood perturbatively. Therefore,
there is no heterotic string theory dual to our model whose conformal
field theory knows about this gauge symmetry factor.

Secondly the counting of moduli, i.e., chargeless hypermultiplets is
curious. This should be equal to the number of deformations of the K3,
plus the number of deformations of the simple instantons still free, minus the
number of hidden obstructers, minus the number of moduli
required to force the two obstructers to meet. The fact that there are
$22+5n-k$ moduli shows that this latter number of moduli, required to be
tuned to make to two obstructers meet, is equal to two.

This tuning must correspond to bending the K3 around as to bring two
orbifold points together in the right way. This should result in a
more complicated orbifold singularity. It looks as if the number of
blow-ups required to smooth this orbifold singularity is equal to two,
from the blow-up modes we already had, plus two more from the tuning
required. This suggests that the resulting quotient singularity is
either of the type $A_4$ (i.e., $\C^2/\Z_5$) or $D_4$ (i.e., $\C^2$
divided by the discrete quaternion group). It would be interesting
to study this further.

If we continue further and attempt to bring three hidden obstructers
together by giving $p_1$ a triple zero, we obtain a total degree at
the collision equal to $(8,12,24)$. After blowing up this point, the
degrees along the exceptional divisor are $(4,6,12)$. While degrees
greater than, or equal to, $(4,6,12)$ are admissible at points within
the discriminant, they are not acceptable along curves. The \CY\
condition is violated if we attempt to blow up. We therefore have
no further extremal transitions associated to three colliding hidden
obstructers. 

%%%%%%%%%%%%%%%%%%%%%%%%%%%%%%%%%%%%%%%%%%%%%%%%%%%%%%%%%%%%%%%%%%%

\section*{Acknowledgements}

I thank M.~Gross for explaining to me much of the technology of elliptic
fibrations used in this paper.
It is also a pleasure to thank O.~Aharony, C.~Johnson, S.~Kachru,
D.~Morrison, J.~Polchinski, N.~Seiberg and E.~Silverstein for useful
conversations. 
The author is supported by DOE grant DE-FG02-96ER40959. 

%%%%%%%%%%%%%%%%%%%%%%%%%%%%%%%%%%%%%%%%%%%%%%%%%%%%%%%%%%%%%%%%%%%

\section*{Appendix}    \label{s:app}

Let $X$ be an algebraic threefold which admits an elliptic fibration,
$p:X\to\Theta$, for some complex surface, $\Theta$. The elliptic
fibres degenerate over the discriminant, $\Delta\subset\Theta$. At a
smooth point in $\Delta$, the bad fibres are classified by the
Weierstrass classification (see, for example, \cite{me:lK3}). In
general however $\Delta$ has singularities, usually formed by
intersections of irreducible components of $\Delta$. In this appendix
we discuss what happens to the bad fibre over such singularities in
$\Delta$.

This problem was studied by Miranda in \cite{Mir:fibr}. 
It has also been analyzed in \cite{BKV:enhg} in terms of Tate's
algorithm. 
We will adopt Miranda's method as it is slightly better
suited to our approach and yields some Euler characteristics which are
required for some points in the main text.
Part of Miranda's approach
was to blow up $\Delta$ until it had only double points. In other
words, he needed only to consider {\em transverse\/} collisions of
two curves within $\Delta$. Such collisions are classified by the
generic fibre type over each of the two curves. In addition some
collisions could be reduced to other types by blowing up the double
point. As such he needed only to consider a subset of all possible
collisions.

Our problem is not quite the same as Miranda's. Blowing up the base,
$\Theta$, will affect the canonical class of $X$, which we want to be
trivial. Sometimes one {\em must\/} blow up the base (as in many
example in the main text) in order to achieve $K_X=0$. In many other
cases blowing up the base would destroy $K_X=0$. We find therefore
that Miranda's classification is not sufficient for us. We must often
deal with collisions within $\Delta$ without blowing them up. As such
there are considerably many more possibilities than Miranda
considered.
See \cite{BJ:collide} for a discussion of some aspects of F-theory
which do fall into Miranda's classification.

Fortunately Miranda's methods did not rely on the assumption that
$\Delta$ contained only double points. Let us review the
construction. Begin with the case of a complex {\em surface\/}, $S$,
which is an elliptic fibration, $\pi:S\to B$, where $B$ is an
algebraic curve. Let $z$ be an affine coordinate in $B$. If this
fibration has a global section then we may write the fibration in
Weierstrass form
\begin{equation}
  y^2 = x^3 + a(z)x + b(z).		\label{eq:Wei}
\end{equation}
The discriminant is then given by $\delta=4a^3+27b^2$.

An elliptic curve may be written as a double cover of $\P^1$ branched
at four points. Indeed, the Weierstrass form exhibits this property
--- $y$ has two solutions for any $x$ except at the roots of the
right hand side of (\ref{eq:Wei}). There are three roots of this cubic
plus one solution ``at infinity''. As such $S$ may be considered as a
double cover of a $\P^1$-bundle over $B$ branched over the curve
$x^3 + a(z)x + b(z)$ and the global section at infinity.

We may draw a typical model for $S$ as 
\begin{equation}
\setlength{\unitlength}{0.008750in}%
\begin{picture}(200,155)(40,645)
\thinlines
\put( 60,660){\vector( 1, 0){180}}
\put( 60,660){\vector( 0, 1){140}}
\multiput(120,800)(0.00000,-5.85366){21}{\line( 0,-1){  2.927}}
\multiput(180,800)(0.00000,-5.85366){21}{\line( 0,-1){  2.927}}
\put( 95,795){
}
\put(100,765){
}
\put(145,645){\makebox(0,0)[lb]{$z$}}
\put( 40,725){\makebox(0,0)[lb]{$x$}}
\put(125,670){\makebox(0,0)[lb]{$\mathrm{I}_0$}}
\put(185,670){\makebox(0,0)[lb]{$\mathrm{I}_1$}}
\end{picture}
\end{equation}
In this graph, the solid lines represent the branch locus (with the
section at infinity at the top) and the two dotted lines represent $\P^1$
fibres for fixed values of $z$. The generic fibre on the left intersects the
branch locus 4 times. The double cover of this is a smooth
elliptic. This is an $\mathrm{I}_0$ fibre. On the fibre on the right,
two of the branch points have coalesced. This amounts to shrinking a
cycle in the elliptic down to a point and, as such, is a curve with a
double point. This is an $\mathrm{I}_1$ fibre. $\delta$ will have a
single zero at this point in $B$. Even though the $\mathrm{I}_1$
fibre is itself singular, $S$ is smooth.

It is possible for the branch locus to degenerate further to produce
higher zeros in $\delta$. As an example let
\begin{equation}
\begin{split}
y^2 = x^3 - 3x + 2+z^N.
\end{split}
\end{equation}
If $N$ is even this looks like
\begin{equation}
\setlength{\unitlength}{0.006250in}%
\begin{picture}(70,140)(80,640)
\thinlines
\multiput(120,780)(0.00000,-8.00000){18}{\line( 0,-1){  4.000}}
\put(105,780){
}
\put( 80,745){
}
\end{picture}
\end{equation}
Now $S$ is singular at $(x,y,z)=(1,0,0)$. We may resolve $S$ by
blowing this point up. We may follow the blow up in our picture. For
example in the case $N=4$ we have
\begin{equation}
\setlength{\unitlength}{0.006250in}%
\begin{picture}(530,159)(80,625)
\thinlines
\multiput(120,780)(0.00000,-8.00000){18}{\line( 0,-1){  4.000}}
\put(105,780){
}
\put( 80,745){
}
\multiput(290,780)(0.00000,-8.00000){18}{\line( 0,-1){  4.000}}
\multiput(270,680)(7.74194,0.00000){16}{\line( 1, 0){  3.871}}
\put(275,780){
}
\put(265,740){
}
\multiput(505,780)(0.00000,-8.00000){18}{\line( 0,-1){  4.000}}
\multiput(485,680)(7.74194,0.00000){16}{\line( 1, 0){  3.871}}
\multiput(580,760)(0.00000,-8.00000){13}{\line( 0,-1){  4.000}}
\put(180,720){\vector( 1, 0){ 60}}
\put(400,720){\vector( 1, 0){ 55}}
\put(490,780){
}
\put(475,735){
}
\put(507,625){\makebox(0,0)[lb]{$f_0$}}
\put(610,675){\makebox(0,0)[lb]{$f_1$}}
\put(580,770){\makebox(0,0)[lb]{$f_2$}}
\end{picture}   
   \label{eq:resI4}
\end{equation}
The new curves may, or may not, be in the branch locus. The rule is
that they are in the branch locus if and only if the total degree of
branch divisor at the point blown up is odd. In the above case this
degree is always two. We denote the fact that the new curves are not
in the branch locus by drawing them as dotted lines.
$S$ will be smooth when the branch locus is smooth and so after these
two blow-ups we are done. The curve $f_0$ is the proper transform of
the original bad fibre. Note that it only intersects the branch locus
twice. Thus, the double cover of this is a rational curve, rather than
an elliptic. The new curve $f_2$ is also branched twice and so will
map to a rational curve in the double cover. $f_1$ is not branched at
all and so must map to {\em two\/} rational curves in the double
cover. The resulting configuration of curves in the double cover is
\begin{equation}
\setlength{\unitlength}{0.006250in}%
\begin{picture}(105,119)(120,700)
\thinlines
\put(140,800){\line( 0,-1){100}}
\put(120,780){\line( 1, 0){100}}
\put(200,800){\line( 0,-1){100}}
\put(120,720){\line( 1, 0){100}}
\put(225,775){\makebox(0,0)[lb]{$f_1$}}
\put(225,715){\makebox(0,0)[lb]{$f_1$}}
\put(200,805){\makebox(0,0)[lb]{$f_2$}}
\put(140,805){\makebox(0,0)[lb]{$f_0$}}
\end{picture}
\end{equation}
This is Kodaira's $\mathrm{I}_4$ fibre.
Subtracting the curve $f_0$, which was already in $S$, we see that the
exceptional divisor within $S$ produced by the blow-up is a chain of
three $\P^1$'s. This is the resolution of the surface singularity
$A_3$ in the usual $A$-$D$-$E$ classification. This method of using a
double cover is probably the best for finding the blow-ups of surface
singularities and may be applied to all of the $A$-$D$-$E$ series.
The type of bad fibres can be classified according to the degree of
vanishing of $a$, $b$, and $\delta$.

Now let us return to our elliptic threefold, $X$. Let $s$ and $t$ be
affine coordinates in the base, $\Theta$. Over a generic point in the
discriminant we may put $z$ equal to a generic linear combination of
$s$ and $t$ and reduce to the elliptic surface case. There is nothing
to stop us putting such a generic slice through a bad point in the
discriminant. The degrees of $(a,b,\delta)$ will jump at such a point.

Consider a transverse intersection of two curves, $D_1$ and $D_2$,
within $\Delta$. The degrees of $(a,b,\delta)$ in our generic
slice given by $z$ will then simply be the sum of the corresponding
degrees along $D_1$ and $D_2$. One may expect them to be higher for
non-transverse intersections however.

For example, let us consider the case given by (\ref{eq:simplek}) of a
curve of $\mathrm{I}_1$ fibres colliding with a curve of $\mathrm{I}_{2k}$
fibres given by
\begin{equation}
\begin{split}
a &= t^k-3s^2\\
b &= -s(t^k-2s^2)\\
\delta &= t^{2k}(4t^k-9s^2).
\end{split}
\end{equation}
The degrees along $4t^k-9s^2=0$ are $(0,0,1)$ (for an $\mathrm{I}_1$ fibre)
and along $t=0$ are $(0,0,2k)$ (for an $\mathrm{I}_{2k}$ fibre). At
$s=t=0$ these curves collide and the total degrees are $(2,3,2k+2)$,
assuming $k\geq2$, (which is an $\Ist{2k-4}$ fibre).

To resolve $X$ we certainly need to begin by blowing up the fibres
along the generic parts of $\Delta$ as in the surface case. Each time
we do a blow-up of the generic points, the fibres at the collisions
will also be partially resolved. In some simple cases the fibres at
the collisions will automatically be fully resolved as usual by this
process but, more usually, we will only end up with a partial
resolution. At this point in the resolution process, $X$ may already
be smooth or it may require a ``small resolution'' at the
collision. Occasionally it cannot be resolved but this will not happen
in any examples here.

Let us follow our example for this process. The $\mathrm{I}_1$ fibres
require no blow-ups so we just have to consider the $\mathrm{I}_{2k}$
blow-up. Let us assume $k=2$ so that the resolution follows the
sequence given in (\ref{eq:resI4}). In this case the partial
resolution of the $\Ist{0}$ fibre at the collision proceeds as
\begin{equation}
\setlength{\unitlength}{0.006250in}%
\begin{picture}(520,159)(90,625)
\thinlines
\multiput(505,780)(0.00000,-8.00000){18}{\line( 0,-1){  4.000}}
\multiput(580,760)(0.00000,-8.00000){13}{\line( 0,-1){  4.000}}
\put(180,720){\vector( 1, 0){ 60}}
\put(400,720){\vector( 1, 0){ 55}}
\multiput(120,780)(0.00000,-8.00000){18}{\line( 0,-1){  4.000}}
\multiput(290,780)(0.00000,-8.00000){18}{\line( 0,-1){  4.000}}
\put(270,680){\line( 1, 0){120}}
\put(485,680){\line( 1, 0){120}}
\put(490,780){
}
\put(105,780){
}
\put(275,780){
}
\put(100,705){
}
\put(310,710){
}
\put(515,710){
}
\put(505,625){\makebox(0,0)[lb]{$f_0$}}
\put(610,675){\makebox(0,0)[lb]{$f_1$}}
\put(580,770){\makebox(0,0)[lb]{$f_2$}}
\end{picture}
\end{equation}
Note that the first blow-up occurs at a degree 3 point in the branch
locus. The exceptional curve is therefore in the branch locus. Note
that at the end the branch locus is still colliding with itself. If
this were a generic point in the discriminant we would have to
continue the blow-up. In this case however, we are done. $X$ is now
smooth.
The fibre over the collision is the double cover of this which is
given as follows:
\begin{equation}
\setlength{\unitlength}{0.006250in}%
\begin{picture}(105,119)(120,700)
\thinlines
\put(140,800){\line( 0,-1){100}}
\put(200,800){\line( 0,-1){100}}
\put(120,720){\line( 1, 0){100}}
\put(225,715){\makebox(0,0)[lb]{$f_1$}}
\put(200,805){\makebox(0,0)[lb]{$f_2$}}
\put(140,805){\makebox(0,0)[lb]{$f_0$}}
\end{picture}
\end{equation}
This has Euler characteristic 4. For general $k$ the Euler
characteristic is $2+k$.

We are now in a position to read off the monodromy. Note that $f_1$
appears within the branch locus at the collision point and yet there
are two $f_1$ curves in the $\mathrm{I}_4$ fibre away from the
collision. So long as the Weierstrass form gives $y^2$ as a generic
function of $s$ then an orbit around $s=0$ within the $t=0$ line will
exchange the two $f_1$ curves in the fibre. Thus, this particular
collision will induce monodromy. In general this collision will
produce the expected monodromy in $\mathrm{I}_{2k}$ fibres to produce
$\sp(k)$.

One should contrast this with to a {\em transverse\/}
collision of an $\mathrm{I}_{m}$-curve and an $\mathrm{I}_{n}$-curve
in which case there is no monodromy (unless, of course, it's
induced by a collision elsewhere). Analysis in \cite{Mir:fibr} shows
that the Euler characteristic of the fibre over such a collision is $m+n$.
Miranda also considered the case of
an $\mathrm{I}_{2k}$-curve collision with a $\Ist{2m}$-curve which is
relevant for our purposes. In this case the resulting fibre at the
collision point has Euler characteristic $2m+k+6$. Monodromy is
induced on the $\mathrm{I}_{2k}$ fibre but not the $\Ist{2m}$ fibre.

%\bibliographystyle{my-phys}
%\bibliography{string}

\end{document}